\documentclass[twoside,twocolumn,9pt]{article}
\usepackage[super,sort&compress,comma]{natbib}
\usepackage{epsfig}
\usepackage[version=3]{mhchem}
\usepackage[left=1.5cm, right=1.5cm, top=1.785cm, bottom=2.0cm]{geometry}
\usepackage{balance}
\usepackage{mathptmx}
\usepackage{sectsty}
\usepackage{graphicx} 
\usepackage{lastpage}
\usepackage[format=plain,justification=justified,singlelinecheck=false,font={stretch=1.125,small,sf},labelfont=bf,labelsep=space]{caption}
\usepackage{siunitx}
\usepackage{float}
\usepackage{fancyhdr}
\usepackage{fnpos}
\usepackage[scr=esstix,cal=boondox]{mathalfa}
\usepackage[english]{babel}
\addto{\captionsenglish}{%
  
}
\usepackage{array}
\usepackage{droidsans}
\usepackage{charter}
\usepackage[T1]{fontenc}
\usepackage[usenames,dvipsnames]{xcolor}
\usepackage{setspace}
\usepackage[compact]{titlesec}
\usepackage[colorlinks=true,
            linkcolor=blue,
            citecolor=red,
            urlcolor=cyan]{hyperref}
\usepackage{comment}
\usepackage{subfig}
\usepackage{subcaption}
\usepackage{amsfonts}
\usepackage{bm}
\usepackage{upgreek}
\usepackage{cleveref}
\usepackage{lmodern}
\DeclareMathOperator{\divr}{Div}

\newcommand{\Xt}{\bm X_{\rm t}}
\newcommand{\Xb}{\bm X_{\rm b}}

\newcommand{\Ht}{\bm H_{\rm t}}
\newcommand{\Hb}{\bm H_{\rm b}}
\newcommand{\Kt}{\bm K_{\rm t}}
\newcommand{\Kb}{\bm K_{\rm b}}

\newcommand{\phit}{\bm{\phi}_{\rm t}}
\newcommand{\phib}{\bm{\phi}_{\rm b}}
\newcommand{\rt}{\bm r_{\rm t}}
\newcommand{\rb}{\bm r_{\rm b}}

\usepackage{xcolor}
\definecolor{C0}{HTML}{1F77B4}
\definecolor{C1}{HTML}{FF7F0E}
\definecolor{C2}{HTML}{2ca02c}
\definecolor{C3}{HTML}{d62728}
\definecolor{C4}{HTML}{9467bd}
\definecolor{C5}{HTML}{8c564b}
\def\usetodonotes{} 
\ifdefined\usetodonotes
\usepackage[colorinlistoftodos]{todonotes}

\usepackage{xpatch,environ,ragged2e,setspace}
\RenewEnviron{abstract}{%
  \xdef\theabstracttext{%
    \unexpanded{%
      \def\baselinestretch{2}\noindent\unskip\textbf{Abstract}\par\medskip
      \noindent\unskip\ignorespaces}%
    \unexpanded\expandafter{\BODY}%
  }%
}
\def\theabstracttext{}
\xpatchcmd{\pprintMaketitle}
  {\ifvoid\absbox}
  {\ifx\theabstracttext\empty\else\printtheabstracttext\fi\ifvoid\absbox}
  {}{}
\newcommand{\printtheabstracttext}{{%
  \begin{trivlist}
  \normalfont\normalsize
  \item\relax
  \doublespacing\theabstracttext
  \end{trivlist}
}}
\fi

\usepackage[commandnameprefix=always,commentmarkup=todo,authormarkup=none]{changes}
\definechangesauthor[color=C1]{nikhil}

\usepackage{epstopdf}

\definecolor{cream}{RGB}{222,217,201}

\begin{document}

\pagestyle{fancy}
\thispagestyle{plain}
\fancypagestyle{plain}{
\renewcommand{\headrulewidth}{0pt}
}

\makeFNbottom
\makeatletter
\renewcommand\LARGE{\@setfontsize\LARGE{15pt}{17}}
\renewcommand\Large{\@setfontsize\Large{12pt}{14}}
\renewcommand\large{\@setfontsize\large{10pt}{12}}
\renewcommand\footnotesize{\@setfontsize\footnotesize{7pt}{10}}
\makeatother

\renewcommand{\thefootnote}{\fnsymbol{footnote}}
\renewcommand\footnoterule{\vspace*{1pt}%
\color{cream}\hrule width 3.5in height 0.4pt \color{black}\vspace*{5pt}} 
\setcounter{secnumdepth}{5}

\makeatletter 
\renewcommand\@biblabel[1]{#1}            
\renewcommand\@makefntext[1]%
{\noindent\makebox[0pt][r]{\@thefnmark\,}#1}
\makeatother 
\renewcommand{\figurename}{\small{Fig.}~}
\sectionfont{\sffamily\Large}
\subsectionfont{\normalsize}
\subsubsectionfont{\bf}
\setstretch{1.125} 
\setlength{\skip\footins}{0.8cm}
\setlength{\footnotesep}{0.25cm}
\setlength{\jot}{10pt}
\titlespacing*{\section}{0pt}{4pt}{4pt}
\titlespacing*{\subsection}{0pt}{15pt}{1pt}
\DeclareSIUnit\angstrom{\text{Å}}

\fancyfoot{}
\fancyfoot[LO,RE]{\vspace{-7.1pt}\includegraphics[height=9pt]{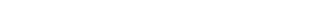}}
\fancyfoot[CO]{\vspace{-7.1pt}\hspace{13.2cm}\includegraphics{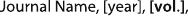}}
\fancyfoot[CE]{\vspace{-7.2pt}\hspace{-14.2cm}\includegraphics{head_foot/RF}}
\fancyfoot[RO]{\footnotesize{\sffamily{1--\pageref{LastPage} ~\textbar  \hspace{2pt}\thepage}}}
\fancyfoot[LE]{\footnotesize{\sffamily{\thepage~\textbar\hspace{3.45cm} 1--\pageref{LastPage}}}}
\fancyhead{}
\renewcommand{\headrulewidth}{0pt} 
\renewcommand{\footrulewidth}{0pt}
\setlength{\arrayrulewidth}{1pt}
\setlength{\columnsep}{6.5mm}
\setlength\bibsep{1pt}

\makeatletter 
\newlength{\figrulesep} 
\setlength{\figrulesep}{0.5\textfloatsep} 

\newcommand{\topfigrule}{\vspace*{-1pt}%
\noindent{\color{cream}\rule[-\figrulesep]{\columnwidth}{1.5pt}} }

\newcommand{\botfigrule}{\vspace*{-2pt}%
\noindent{\color{cream}\rule[\figrulesep]{\columnwidth}{1.5pt}} }
\newcommand{\dblfigrule}{\vspace*{-1pt}%
\noindent{\color{cream}\rule[-\figrulesep]{\textwidth}{1.5pt}} }
\newcommand{\divrt}{\divr}
\newcommand{\divrb}{\divr}
\makeatother

\twocolumn[
  \begin{@twocolumnfalse}
{\includegraphics[height=30pt]{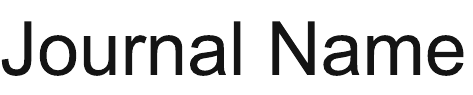}\hfill\raisebox{0pt}[0pt][0pt]{\includegraphics[height=55pt]{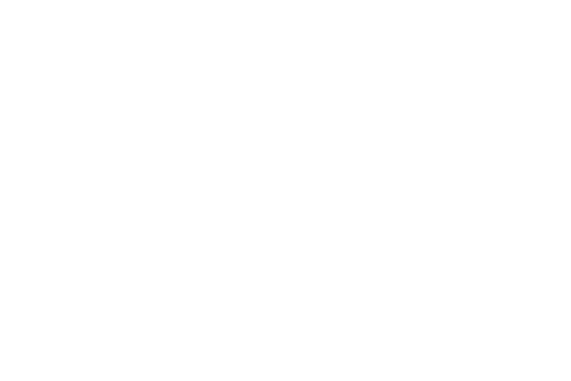}}\\[1ex]
\includegraphics[width=18.5cm]{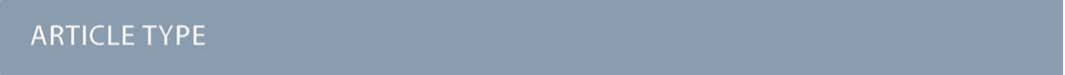}}\par
\vspace{1em}
\sffamily
\begin{tabular}{m{4.5cm} p{13.5cm} }

ad\includegraphics{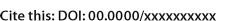} & \noindent\LARGE{\textbf{Multiscale analysis of large twist ferroelectricity and swirling dislocations in bilayer hexagonal boron nitride}} \\
\vspace{0.3cm} & \vspace{0.3cm} \\

 & \noindent\large{Md Tusher Ahmed,\textit{$^{a\ddag}$} Chenhaoyue Wang,\textit{$^{b\ddag}$} Amartya S. Banerjee,\textit{$^{b}$} and Nikhil Chandra Admal\textit{$^{a}$}$^{\ast}$} \\

\includegraphics{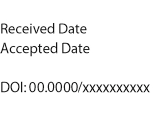} & \noindent\normalsize{
With its atomically thin structure and intrinsic ferroelectric properties, heterodeformed bilayer hexagonal boron nitride (hBN) has gained prominence in next-generation non-volatile memory applications. However, studies to date have focused almost exclusively on small$-$twist bilayer hBN, leaving the question of whether ferroelectricity can persist under small heterostrain and large heterodeformation entirely unexplored. In this work, we establish the crystallographic origin of ferroelectricity in bilayer hBN configurations heterodeformed relative to high-symmetry configurations such as the AA-stacking and the  $21.786789^\circ$ twisted configuration ($\Sigma 7$), using Smith normal form bicrystallography.  We then demonstrate out-of-plane ferroelectricity in bilayer hBN across configurations vicinal to both the AA and $\Sigma7$ stackings. Atomistic simulations reveal that AA-vicinal systems support ferroelectricity under both small twist and small strain, with polarization switching in the latter governed by the deformation of swirling dislocations rather than the straight interface dislocations seen in the former. For $\Sigma7$-vicinal systems, where existing interatomic potentials underperform particularly under extreme out-of-plane compression, we develop a density-functional-theory-informed continuum framework—the bicrystallography-informed frame-invariant multiscale (BFIM) model, which captures out-of-plane ferroelectricity in heterodeformed configurations vicinal to the $\Sigma 7$ stacking.
Interface dislocations in these large heterodeformed bilayer configurations exhibit markedly smaller Burgers vectors compared to the interface dislocations in small-twist and small-strain bilayer hBN. The BFIM model reproduces experimental results and provides a powerful, computationally efficient framework for predicting ferroelectricity in large-unit-cell heterostructures where atomistic simulations are prohibitively expensive.
}

\end{tabular}

 \end{@twocolumnfalse} \vspace{0.6cm}

  ]

\renewcommand*\rmdefault{bch}\normalfont\upshape
\rmfamily
\section*{}
\vspace{-1cm}


\footnotetext{\textit{$^{a}$~Department of Mechanical Science and Engineering, University of Illinois at Urbana-Champaign, Urbana, IL, USA}}

\footnotetext{\textit{$^{b}$~Department of Materials Science and Engineering, University of California, Los Angeles., Los Angeles, CA, USA}}


\footnotetext{\ddag~These authors contributed equally to this work}

\footnotetext{$^{\ast}$~Corresponding author email: admal@illinois.edu}


\section{Introduction}
Since the discovery of ferroelectricity in Rochelle salt by \citet{valasek1921piezo}, ferroelectrics have emerged as an important class of materials for the development of non-volatile memory devices \cite{mikolajick2018ferroelectric,zhao2020ultrathin}. Ferroelectricity, the property of reversing the spontaneous polarization of certain materials through the application of an electric field, enables instantaneous reading/writing operations through switching polarizations \cite{muller2012ferroelectricity}. Due to their intrinsic ferroelectricity \cite{fei2018ferroelectric,yasuda2021stacking}, strong resistance to the formation of depolarization field \cite{mehta1973depolarization}, and atomically thin nature, van der Waals (vdW) homo and heterostructures --- e.g., bilayer hexagonal boron nitride, bilayer molybdenum disulfide, and molybdenum disulfide-tungsten disulfide --- have been recognized as promising ferroelectric materials. In particular, the tunability of ferroelectricity through controlled spatially varying relative sliding --- obtained by imposing a relative twist and/or relative strain (heterostrain) between the two layers ---  makes vdW structures suitable for versatile applications in nano- and micro- electronic devices \cite{enaldiev2024dislocations,weston2022interfacial}. We use \emph{heterodeformation} as an umbrella term to refer to relative twist, strain, or a combination of twist and strain to the bilayers.

Ferroelectricity in heterodeformed bilayer 2D materials arises from manipulating triangular domains formed during structural reconstruction of van der Waals (vdW) structures \cite{weston2020atomic, yasuda2021stacking}. These structures consist of two layers with identical or distinct lattices, each containing basis atoms with different polarities, leading to unique polarizations for different vertical stackings \cite{liu2016room, fei2018ferroelectric}. Small twists and/or strains in bilayer vdW structures undergo structural reconstruction mediated by interface dislocations, forming triangular domains with alternating polarizations \cite{enaldiev2021piezoelectric,bennett2023polar} \footnote{Disregistry in twisted bilayer graphene increases significantly as the relative twist angle is increased beyond $2.5^\circ$ \cite{Annevelink_2020}causing overlap between neighbouring dislocations. Thus, we refer to twists greater than $2.5^\circ$ as large twist heterodeformations.}. These domain shapes can be modified via interfacial sliding and bending of the interface dislocations under an electric field, creating a net polarization—a phenomenon known as sliding ferroelectricity \cite{yasuda2021stacking,weston2022interfacial}.  However, current studies of sliding ferroelectricity are mostly limited to twisted bilayer 2D materials with a twist angle, $\theta< 2^\circ$ \cite{woods2021charge,weston2022interfacial}. Tuning of ferroelectricity through the application of heterostrain has not been extensively explored, which motivates us to explore different heterodeformations for which ferroelectricity can be observed. In this study, we consider bilayer hexagonal boron nitride (hBN) as the representative bilayer 2D material due to its analogy to graphene \cite{yang2015graphene}, its superior electric, chemical, and thermal properties \cite{yin2016boron}, and its capability of showing sliding ferroelectricity at high operating temperatures \cite{yasuda2021stacking,woods2021charge}.

Large-scale exploration of ferroelectricity across all possible heterodeformed bilayer hBN arrangements is experimentally impractical, thus motivating alternate avenues for study. Atomic-scale simulations can be very useful for such purposes, though such explorations have two limitations. First, an enormous periodic simulation domain must be designed to remove edge effects and consistently predict ferroelectricity (in the large-body limit) in a heterodeformed bilayer hBN. The required domain size can be very large even for small heterodeformations (e.g., small twist angles) \cite{uchida2014atomic}. Predicting structural reconstruction in such large domains is computationally expensive. Consequently, prior studies have largely focused on small heterodeformations, as systematic exploration of the full heterodeformation space remains prohibitively expensive using atomistic simulations. Second, the predictability of accurate ferroelectric behavior in heterodeformed bilayer hBN requires a reliable interatomic potential to predict consistent structural reconstruction results across every heterodeformation. While structural reconstruction for small-heterodeformed bilayer hBN can be accurately predicted using atomic-scale simulations, the same can not be achieved for large heterodeformed bilayer hBN \footnote{In a previous work \citep{ahmed2024bicrystallography}, we showed that structural relaxation also occurs under small heterodeformations relative to the $\SI{21.786789}{\degree}$ twisted configuration, which is defined as large heterodeformed bilayer homostructures. Interestingly, in such cases, the Burgers vector magnitude is markedly different from the magnitude of the Burgers vector observed in small heterodeformed bilayer homostructures.}. Here, we aim to develop a bicrystallography-informed frame-invariant multiscale (BFIM) model for predicting structural reconstruction in both small and large heterodeformed bilayer hBN under the presence and absence of applied out-of-plane electric fields. While previously available multiscale models are developed for the exploration of ferroelectricity in small heterodeformed bilayer 2D materials only \cite{enaldiev2021piezoelectric,enaldiev2024dislocations}, the BFIM model is unique in its potential to capture ferroelectricity in large heterodeformed bilayer hBN.

The hypothesis of observation of ferroelectricity in large heterodeformed bilayer 2D materials originates from the work of \citet{ahmed2024bicrystallography} who showed that structural reconstruction through the formation of interface dislocations is not limited to small heterodeformation in bilayer 2D materials. In this study, we show through density functional theory (DFT) simulations that two degenerate energy minima can be observed in the generalized stacking fault energy plot of $21.786789^\circ$ twisted bilayer hBN, similar to $0^\circ$ bilayer hBN,  which is not observable through atomic-scale calculations using existing interatomic potentials. Moreover, we show that two energy minima corresponding to the $21.786789^\circ$ twisted bilayer hBN demonstrate alternating out-of-plane polarization similar to the $0^\circ$ bilayer hBN, which is an indicator of the sliding ferroelectricity in large heterodeformed bilayer hBN. Following this, we employ quantum-scale information of stacking energy and polarization to develop the BFIM model for efficient and effective prediction of ferroelectricity of an arbitrary heterodeformed bilayer hBN.         

The remainder of this paper is organized as follows. Section \ref{sec:atomistics} begins with a review of structural reconstruction and the formation of interface dislocations using molecular dynamics (MD) simulations of bilayer hBN subjected to small heterodeformations. Subsequently, MD simulations of GSFE are used to identify the translational invariances in $0^\circ$ and $21.786789^\circ$ twisted bilayer hBN, showing the unreliability of existing interatomic potentials to predict ferroelectricity in such configurations.
In section \ref{sec:cont_fram}, we present the BFIM model for predicting structural reconstruction in heterodeformed bilayer hBN under the presence and absence of applied electric fields. In section \ref{sec:results}, we compare the computed results for small-heterodeformed bilayer hBN using the BFIM model with the results from atomistic simulations using LAMMPS. Following this, we show the ferroelectric domain formation in large heterodeformed bilayer hBN using the BFIM model. We summarize and conclude in section \ref{sec:conclusions}. 
\\

\noindent
\emph{Notation}:
Lowercase bold letters are used to denote vectors, while uppercase bold letters represent second-order tensors unless stated otherwise. The gradient and divergence operators are denoted by the symbols $\nabla$ and $\divr$, respectively. We use the symbol $\cdot$ to denote the inner product of two vectors or tensors. 

\section{Atomistic investigation of ferroelectricity in bilayer hBN: role of interface dislocations}
\label{sec:atomistics}
Ferroelectricity refers to the reversal of spontaneous polarization in certain materials when an external electric field is applied.  In bilayer hBN subjected to a small heterodeformation, ferroelectricity arises due to the presence of AB and BA domains with opposite polarizations normal to the interface. The domains are separated by interface dislocation lines \cite{weston2022interfacial,bennett2023polar,enaldiev2024dislocations}, which form as a result of atomic reconstruction in 2D homo/heterostructures subjected to small heterodeformation. The ferroelectric response of heterodeformed hBN results from the expansion and contraction of the domains when subjected to an out-of-plane electric field. Since the evolution of the domains is driven by the motion of interface dislocations, it is essential to understand the properties of these dislocations to quantify ferroelectricity in heterodeformed bilayer hBN.

\begin{figure}[t]
    \centering
    \includegraphics[width=1\linewidth]{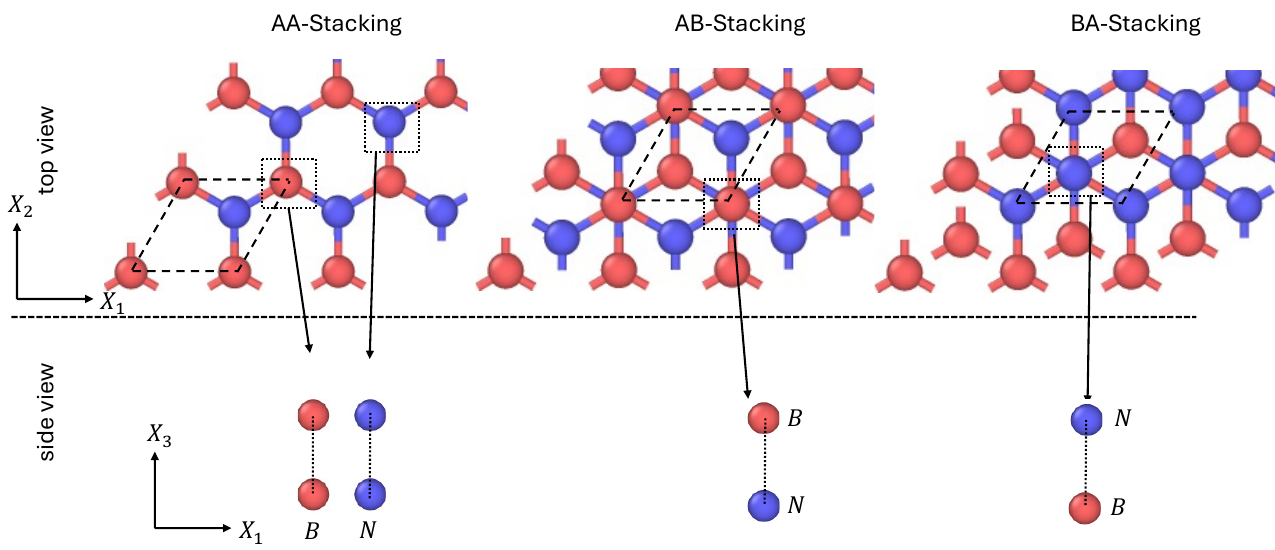}
    \caption{Three characteristics stacking configuration in bilayer hBN. The parallelepiped marked by dotted lines identifies the smallest unit cell.}
    \label{fig:stack_schem}
\end{figure}

This section aims to explore the role of interface dislocations in mediating ferroelectricity through atomistic simulations.
We begin by examining the structure of interface dislocations in bilayer hBN under small relative twists and strains. Then, we demonstrate ferroelectricity by applying an electric field, which causes the dislocation to move, and leads to the expansion and contraction of the AB and BA domains. Finally, we provide arguments supporting the observation of ferroelectricity in bilayer hBN beyond the small heterodeformation range.
We use Large-scale Atomic/Molecular Massively Parallel Simulator (LAMMPS) \cite{thompson2022lammps} to model a heterodeformed bilayer hBN at the atomic scale. The top hBN layer, represented by lattice $\mathcal A$, is constructed using the structure matrix:
\begin{equation*}
    \bm A = \frac{a}{2} \begin{bmatrix}
    0 &  -\sqrt{3}\\
    2 & -1
    \label{eq:A_matrix}
    \end{bmatrix},
\end{equation*}
where the columns of $\bm A$ represent the basis vectors, and $a=2.51~\si{\angstrom}$ is the lattice constant of strain-free hBN. The two basis atoms are positioned at coordinates $\left(0,0\right)$ and $\left(\frac{1}{3},\frac{2}{3}\right)$ relative to the basis vectors of $\mathcal A$. The bottom layer, represented by lattice $\mathcal B$, is constructed using the structure matrix $\bm B=\bm F \bm A$, where $\bm F$ is the heterodeformation gradient. The bilayer is in an \emph{AA stacking} if $\bm F=\bm I$, and relative translations between the two layers will lead to the AB and BA stackings, as shown in \Cref{fig:stack_schem}.

To eliminate edge effects in all simulations, periodic boundary conditions (PBCs) are enforced in the plane of the bilayer. To impose PBCs, the simulation box size must be carefully selected so that the box vectors belong to the deformed lattices of the top and the bottom layers. We use the Smith normal form (SNF) bicrystallography framework developed by \citet{ADMAL2022} to calculate the box vectors. Additionally, SNF bicrystallography describes the translational symmetry of an interface, which will be used in \Cref{sec:cont_fram} to explore ferroelectricity in hBN under large heterodeformations.

The intralayer nearest-neighbor interaction between boron and nitrogen atoms (B-N bond) is modeled using the modified Tersoff potential  \cite{tersoff1988new,mandelli2019princess,ouyang2019mechanical}. The interlayer van der Waals interaction is modeled using the registry-dependent interlayer potential (ILP) potential \cite{ouyang2018nanoserpents} with a cutoff radius of $16~\si{\angstrom}$ to ensure adequate sampling of dispersive interactions. Hexagonal boron nitride consists of boron and nitrogen atoms, which possess average partial charges of $+0.4$ and $-0.4$, respectively, making hBN a polar material \cite{won2008structure}. To account for electrostatics, we included a Coulomb potential with a cutoff of $16~\si{\angstrom}$, consistent with the ILP cutoff. To simulate substrate effects in experiments, a continuum substrate is placed to interact with the hBN layers, as opposed to conventional free-suspended or out-of-plane constrained boundary conditions along the $X_3$ direction \cite{song2018robust,kazmierczak2021strain}. The implementation of the continuum substrate is described in Section S$-1$ of the Supporting Information. 
Atomic reconstruction is simulated by minimizing the total energy using the fast inertial relaxation engine (FIRE) algorithm \citep{bitzek2006fire} with an energy tolerance and force tolerance of $\SI{1e-20}{\eV}$ and $\SI{1e-20}{\eV \per \angstrom}$, respectively. The resulting atomic displacements are analyzed to interpret them in terms of interface dislocations. To calculate the polarization of the system, we employ LAMMPS's built-in dipole moment calculator. Due to the non-uniqueness associated with the absolute polarization of periodic systems, we ultimately measure changes in polarization. In the next section, we discuss the characteristics of interlayer dislocations in bilayer hBN under small heterodeformations. 

\subsection{Characterization of interface dislocations in heterodeformed bilayer hBN}
\label{sec:characterize_atom}
The structure of interface dislocations in hBN varies significantly depending on the relative deformation between the two layers of hBN. To demonstrate this, we conduct structural reconstruction simulations under the following two heterodeformations:
a) a $0.2992634^\circ$ twist, and b) a pure stretch of
\begin{equation}
    \bm U = 
    \begin{bmatrix}
        1.004219 &  0\\
        0        &  1.004219
    \end{bmatrix},
    \label{eqn:U_smallTwist}
\end{equation}
relative to the $0^\circ$ twisted AA-stacked bilayer hBN. The corresponding periodic simulation box vectors, computed from SNF bicrystallography, are shown below
\begin{subequations}
    \label{eqn:pbc_small_strain}
    \begin{align}
        \text{twist: } \bm b_1 &= (480.394592 \, \bm e_1)\si{\angstrom}, 
        \\ \bm b_2 &=
        (240.197294 \,  \bm e_1 +416.033919\, \bm e_2)\si{\angstrom}; \\
        \text{equi-biaxial strain: } \bm b_1 &= (593.619081 \, \bm e_1)\si{\angstrom},
        \\ \bm b_2 &=(293.036378\,  \bm e_1 + 516.248841\, \bm e_2)\si{\angstrom},
    \end{align}
\end{subequations}
where $\bm e_i$s denote the unit vectors parallel to the global axes $X_1$, $X_2$, and $X_3$. The PBCs ensure that the average heterodeformation stays constant during structural relaxation.

\begin{figure}[t]
    \centering
    \subfloat[]
    {
        \centering
        \includegraphics[width=0.5\textwidth]{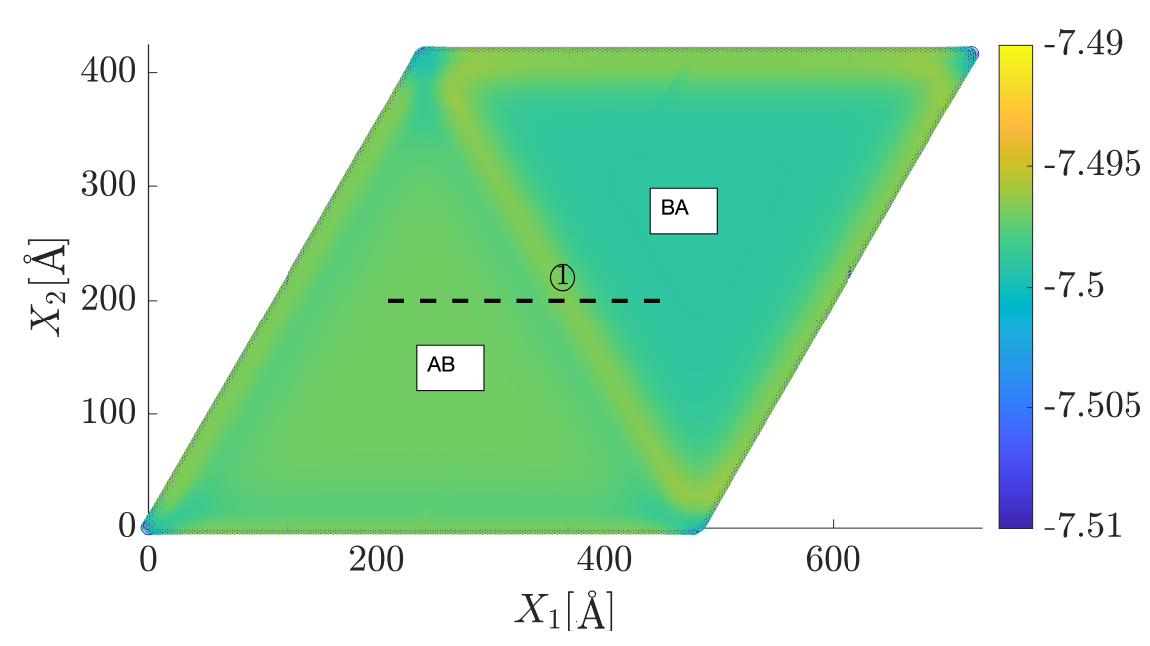}
        \label{fig:twist_relax}
    }\\
    \subfloat[]
    {
        \includegraphics[width=0.43\textwidth]{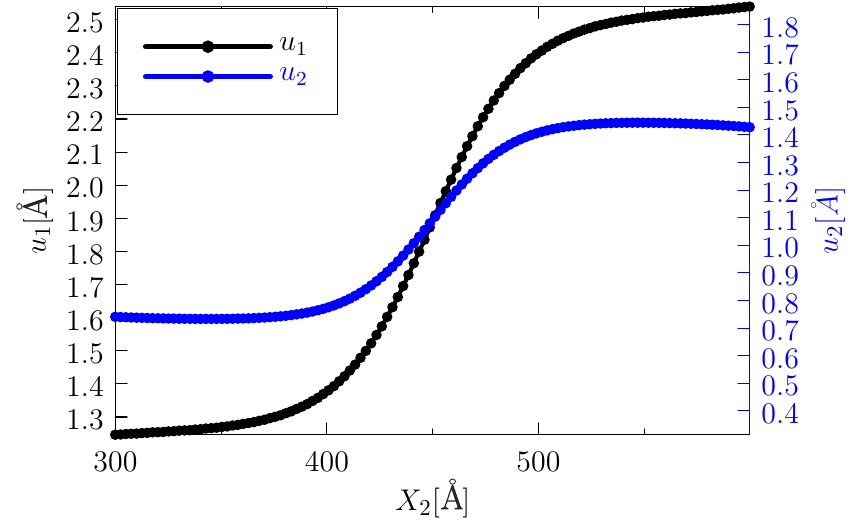}
        \label{fig:Burgers_twist}
    }
    \caption{Atomic reconstruction in a $\SI{0.299}{\degree}$ twisted bilayer hBN. \protect\subref{fig:twist_relax} Atomic energy per atom [in $\si{\eV}$] plot showing a triangular network of interface dislocations that separate the low-energy Bernal stackings. \protect\subref{fig:Burgers_twist} Displacement components $u_1$ and $u_2$ [in $\si{\angstrom}$], measured along the line \textcircled{1} (in \protect\subref{fig:twist_relax}) and relative to the untwisted (AA stacking) configuration, signify the screw characteristic of dislocations in a twisted bilayer hBN.}
    \label{fig:relax_twist}
\end{figure}

The atomic energy map of the relaxed $0.299263^\circ$ twisted bilayer hBN in \Cref{fig:twist_relax} shows the formation of two isoenergetic AB and BA domains, which are separated by interlayer dislocations. \Cref{fig:Burgers_twist} shows a line scan of the displacement components along segment \textcircled{1}. The jumps in displacement components $u_1$ and $u_2$ in \Cref{fig:Burgers_twist} suggest the Burgers vector (jump in the displacement vector) is parallel to the dislocation line. In other words, the interface dislocations in a twisted bilayer hBN have a screw character as supported by prior studies \cite{harley_disloc,Annevelink_2020}.
 
\begin{figure}[t]
    \centering
    \subfloat[]
    {
        \includegraphics[width=0.5\textwidth]{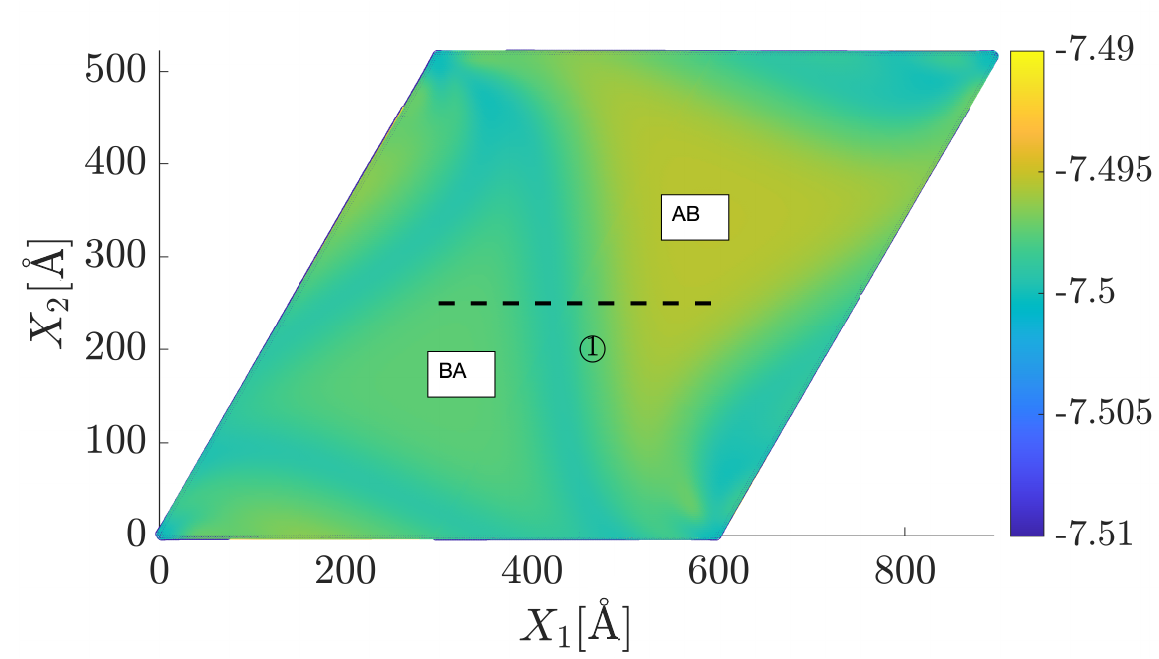}
        \label{fig:strain_relax}
    }\\
    \subfloat[]
    {
        \includegraphics[width=0.43\textwidth]{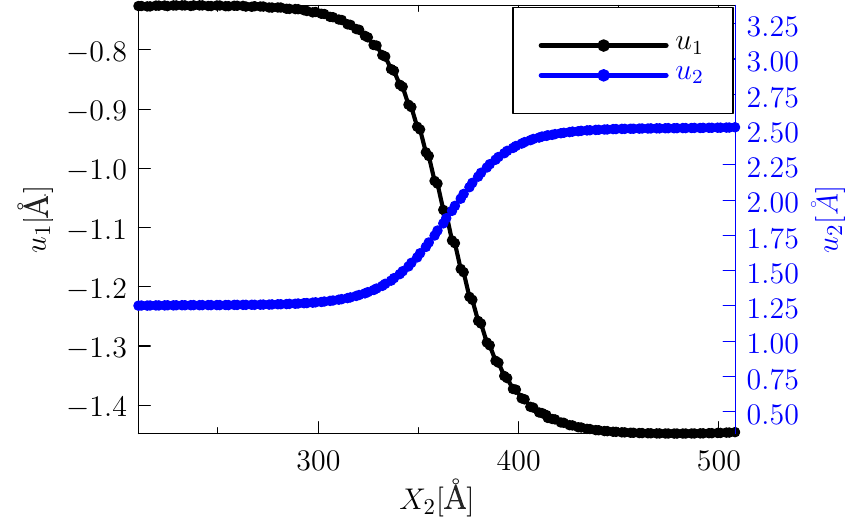}
        \label{fig:Burgers_strain}
    }
    \caption{Atomic reconstruction in a $0.422\%$ equi-biaxial heterostrained bilayer hBN. \protect\subref{fig:strain_relax} Atomic energy per atom [in $\si{\eV}$] plot showing a triangular network of swirling dislocations that separate the low-energy Bernal stackings. \protect\subref{fig:Burgers_strain} Displacement components $u_1$ and $u_2$ [in $\si{\angstrom}$], measured along line \textcircled{1}(in \protect\subref{fig:strain_relax}) and relative to the undeformed (AA stacking) configuration, signify the mixed characteristic of dislocations under equi-biaxial heterostrains.}
    \label{fig:relax_strain}
\end{figure}

On the other hand, a $0.4219\%$ equi-biaxial heterostrained bilayer hBN forms a spiral triangular network of dislocation lines, as illustrated in Figure \ref{fig:strain_relax}. Topologically, a triangular network of straight edge dislocations is consistent with the incompatibility associated with an equi-biaxial heterostrain. However, the dislocation lines twist by $60^\circ$ at the AA junctions, resulting in a swirling network. This swirling occurs because the pure edge dislocations transform locally near the AA junctions to enhance their screw character, thereby releasing the excess strain energy \cite{mesple2023giant,ahmed2025quantifying}. 

\begin{figure}[t]
    \centering
    \subfloat[]
    {
    \includegraphics[width=0.45\textwidth]{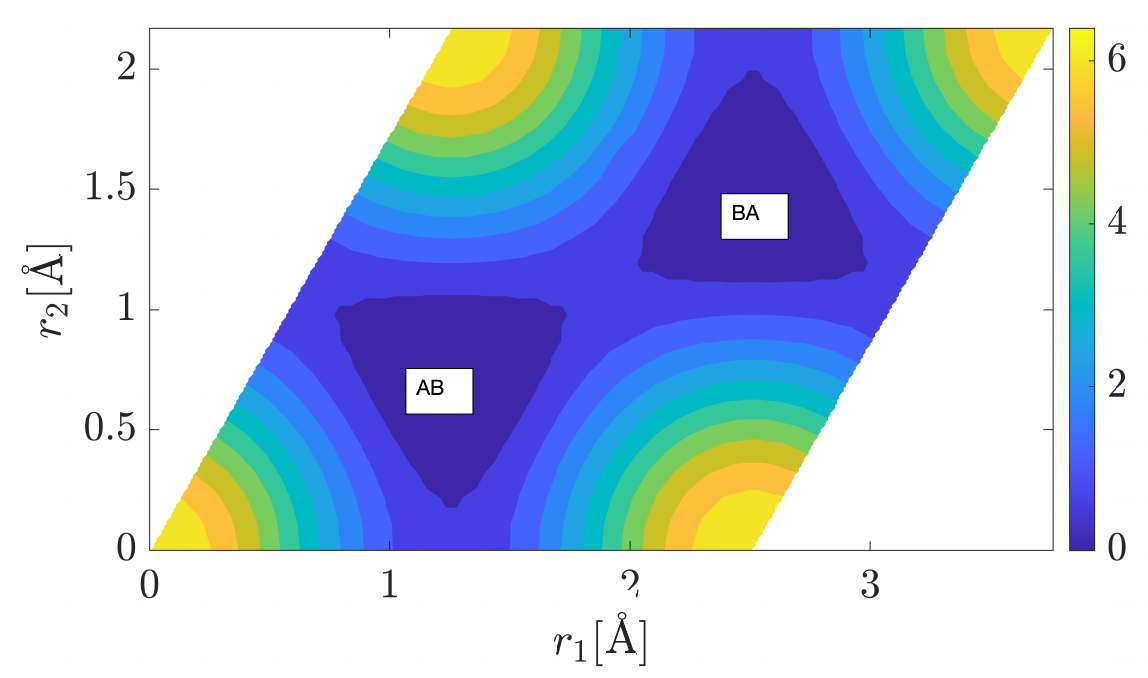}
        \label{fig:gsfe_dft_small}
    }\\
    \subfloat[]
    {
\includegraphics[width=0.45\textwidth]{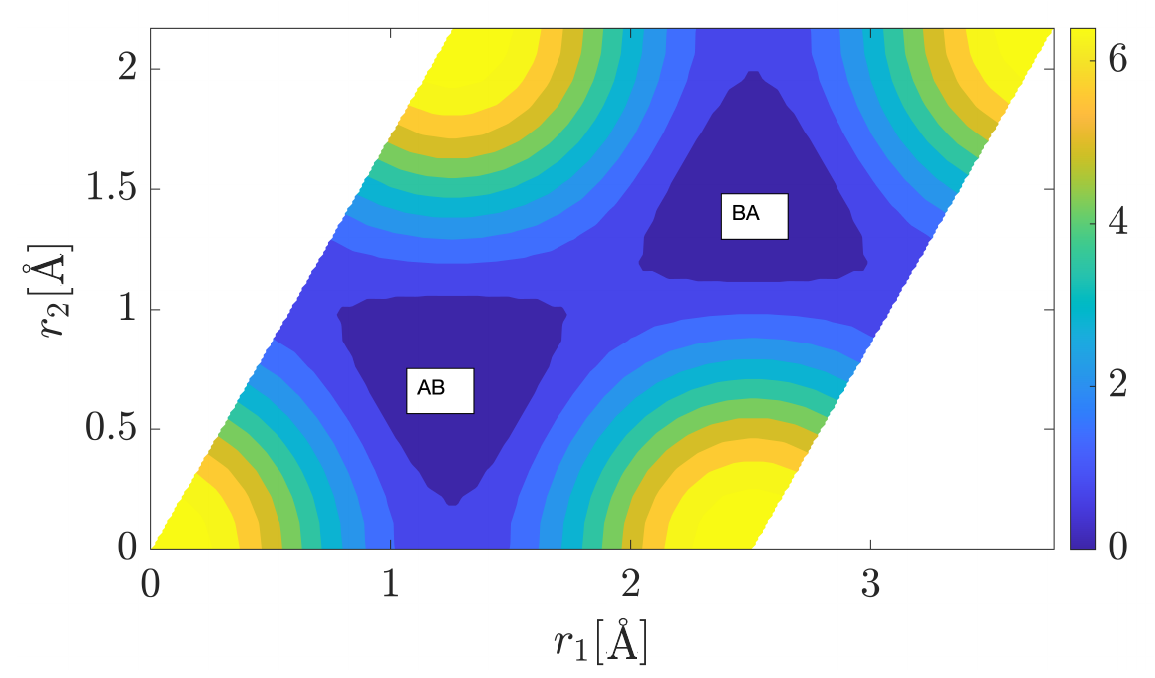}
    \label{fig:gsfe_atom}
    } 
    \caption{Generalized stacking fault energy [$meV\si{\angstrom}^{-2}$] of $0^\circ$-twist bilayer hBN computed using \protect\subref{fig:gsfe_dft_small} DFT \protect\subref{fig:gsfe_atom} LAMMPS, and plotted as functions of the relative displacement between the two layers. The four corners correspond to the AA stacking. 
    }
    \label{fig:gsfe_small_dft_atom}
\end{figure}

Since the formation of AB and BA domains during structural relaxation results from local relative translations between the layers, the generalized stacking fault energy (GSFE) map, which describes the interfacial energy as a function of relative translations, encodes a bilayer's dislocation properties. In particular, the periodicity of GSFE conveys the translational invariance of an interface and defines the sets of interlayer dislocations an interface can host \cite{ADMAL2022,ahmed2024bicrystallography}. 
\Cref{fig:gsfe_small_dft_atom}  shows the GSFE of a $0^\circ$-twist bilayer hBN calculated using density functional theory (DFT), with details in Section S$-3$ of the Supporting Information, and LAMMPS. The agreement between \Cref{fig:gsfe_dft_small,fig:gsfe_atom} confirms the accuracy of the interatomic potential used in our atomistic simulations.
From the figure, we note that three other equivalent low-energy stackings surround each low-energy stacking. This feature leads to the formation of a triangular dislocation network in bilayer hBN. Additionally, the shortest distance between two degenerate minima determines the Burgers vector of an interlayer dislocation, which matches the Burgers vector calculated from the plots in \Cref{fig:Burgers_twist} and \Cref{fig:Burgers_strain}. 
While the GSFE encodes the Burgers vectors of interface dislocations, it cannot describe their response to an external electric field. In the next section, we demonstrate ferroelectricity under small heterodeformations and introduce the polarization map to explain how dislocations respond to an electric field during ferroelectric transitions. 
\subsection{Ferroelectricity in bilayer hBN under small heterodeformations}
\label{sec:small}
In this section, we simulate the ferroelectric transition in a small-heterodeformed bilayer hBN and reveal its crystallographic origin via the polarization landscape. From the previous section, recall that the equi-sized AB and BA triangular domains formed during the structural relaxation of a small-heterodeformed bilayer hBN are energetically equivalent. However, their response to an applied electric field varies due to their polarity. In the AB stacking, boron (positively polarized) sits above nitrogen (negatively polarized), whereas in the BA stacking, the sequence is reversed. This polarity difference results in opposite reactions when an electric field is applied perpendicular to the layers. 

\begin{figure}[t]
    \centering
    \subfloat[]
    {
        \includegraphics[width=0.45\textwidth]{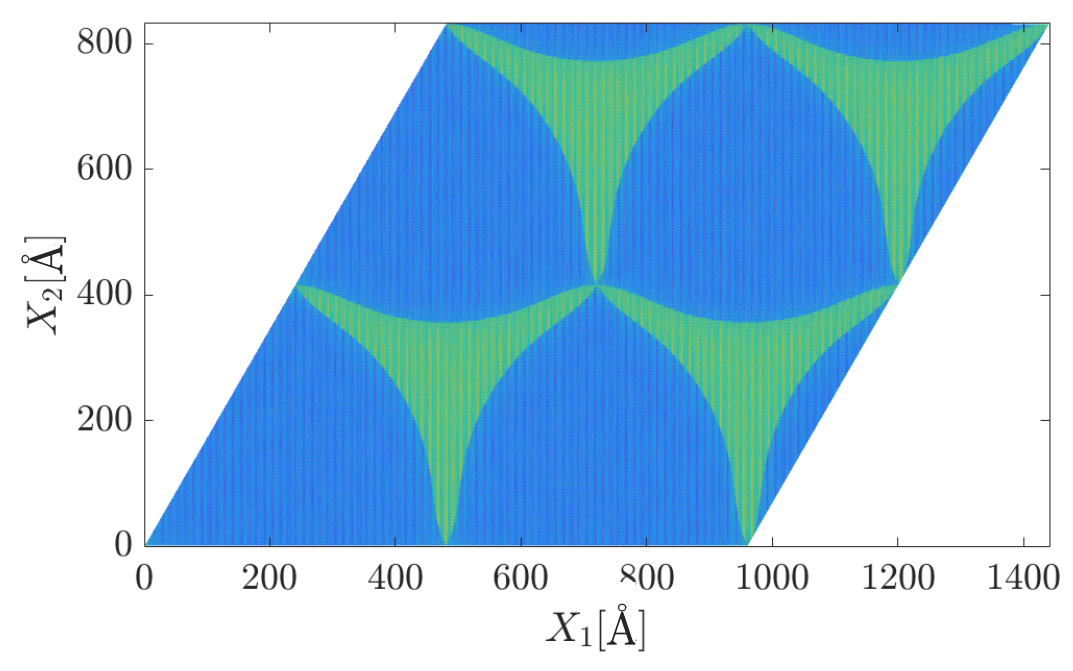}
        \label{fig:twist_pos}
    }\\
    \subfloat[]
    {
        \includegraphics[width=0.45\textwidth]{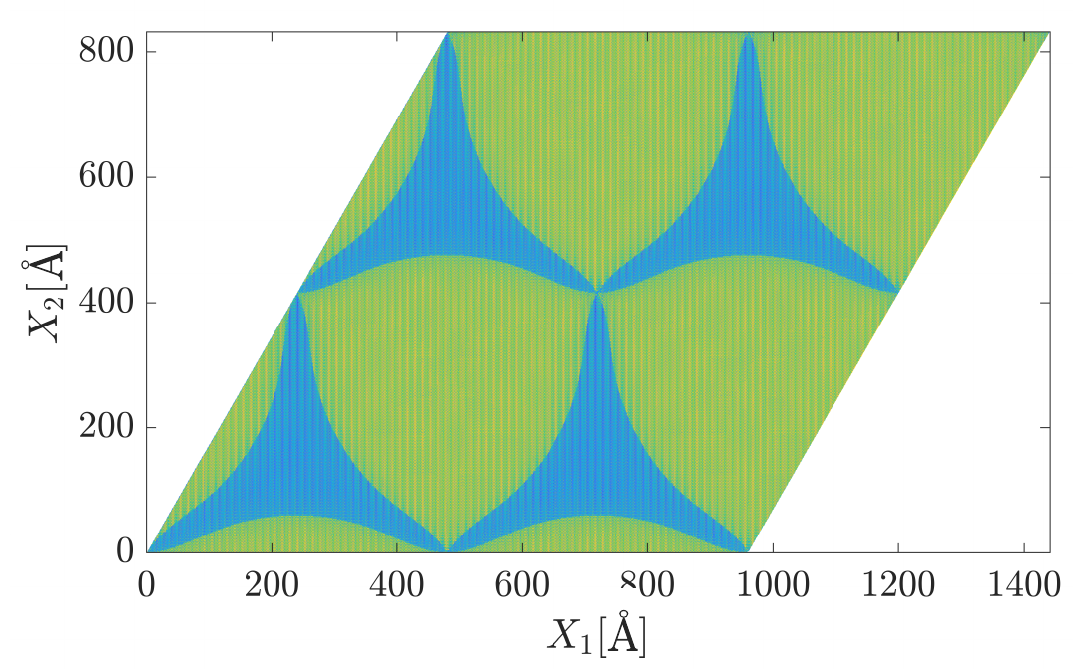}
        \label{fig:twist_neg}
    } 
    \caption{The AB (blue) and BA (green) domains in a relaxed $0.299^\circ$ twisted bilayer hBN subjected to a $\SI{5}{\volt\per\angstrom}$ electric field in the \protect\subref{fig:twist_pos} $+X_3$ direction and \protect\subref{fig:twist_neg} $-X_3$ direction.}
    \label{fig:ferro_atom_twist}
\end{figure}
To illustrate ferroelectricity in a small twist bilayer hBN, we subject the $0.299263^\circ$ twisted hBN to an external electric field of $\SI{5}{\volt\per\angstrom}$ in $\pm X_3$ direction. An electric field $\bm E$ is enforced using the \texttt{fix efield} command in LAMMPS, resulting in force $q\bm{E}$ on an atoms with fixed charge $q$. The electric field induces a transformation in the bilayer structure, as shown in \Cref{fig:ferro_atom_twist}. The blue regions in the figure represent the AB stacking, and the green regions indicate BA stacking. Under a positive electric field, the AB domains expand while the BA domains shrink (\Cref{fig:twist_pos}). The bending of the interlayer dislocations mediates the transformation from one stacking to another.
\begin{figure}[t]
    \centering
    \subfloat[]
    {
        \includegraphics[width=0.45\textwidth]{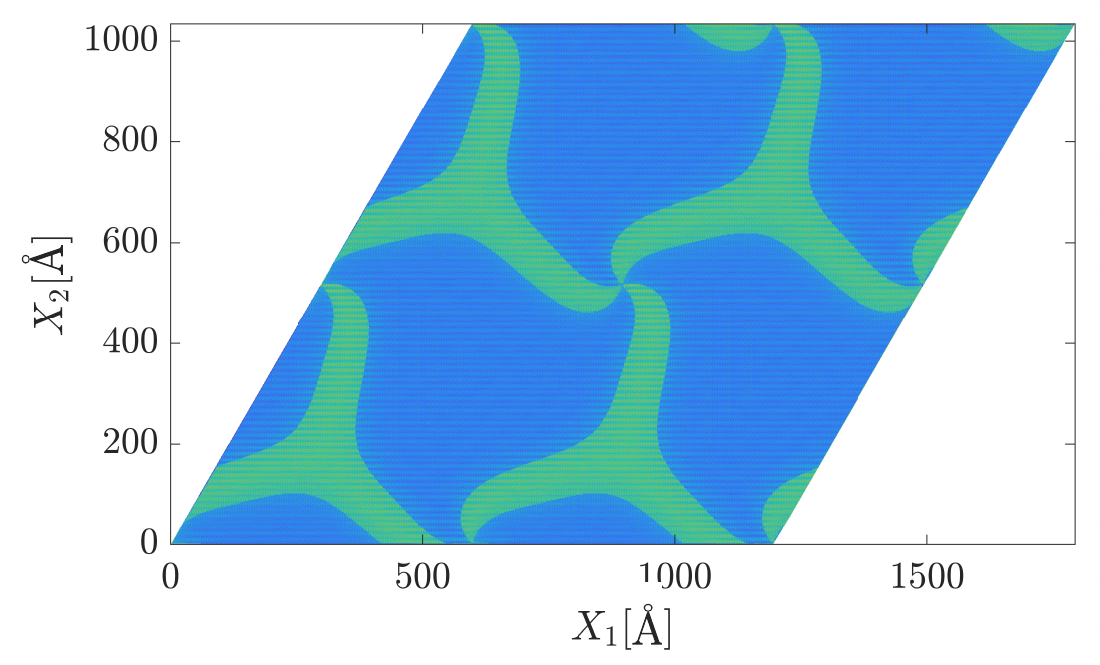}
        \label{fig:str_pos}
    }\\
    \subfloat[]
    {
        \includegraphics[width=0.45\textwidth]{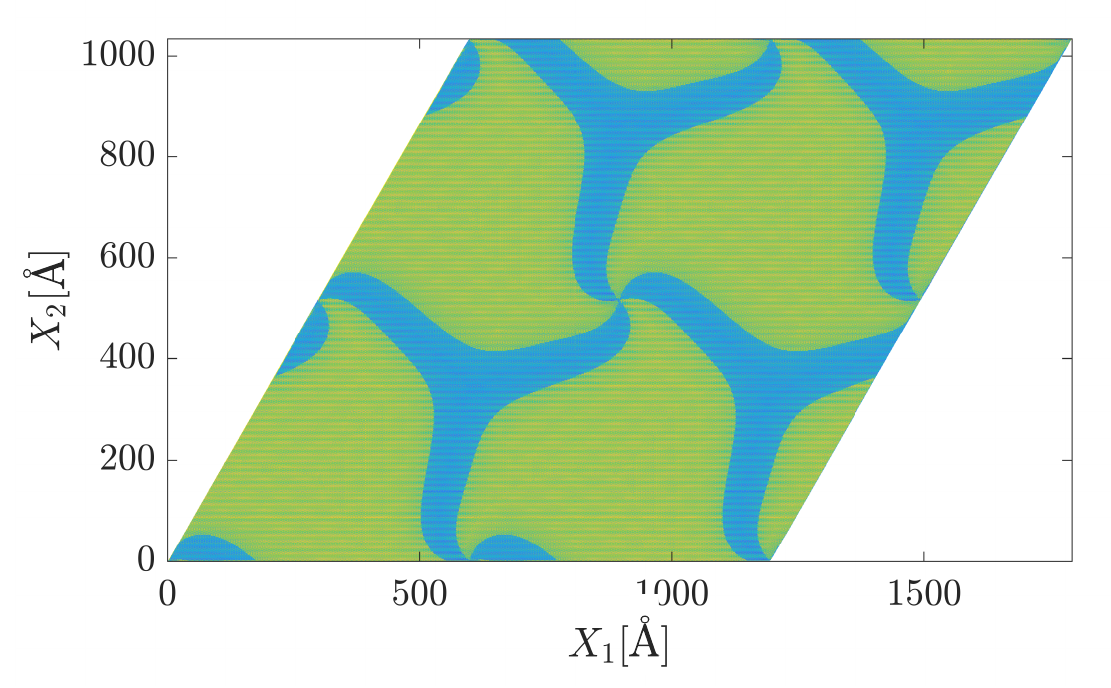}
        \label{fig:str_neg}
    } 
    \caption{Distribution of AB and BA regions in $0.422\%$ equi-biaxially heterostrained bilayer hBN while subjected to applied electric field. \protect\subref{fig:str_pos} and \protect\subref{fig:str_neg} correspond to the relaxed configurations while subjected to positive and negative out-of-plane electric fields, respectively.}
    \label{fig:ferro_atom_str}
\end{figure}
Next, we examine ferroelectricity in a relaxed, equi-biaxially heterostrained hBN. Recall from \Cref{sec:characterize_atom} that the equi-biaxial heterostrain causes spiral dislocations. Similar to the small-twist case, the AB and BA domain areas change when an external electric field is applied, as shown in \Cref{fig:ferro_atom_str}. However, under the out-of-plane electric field of $\SI{5}{\volt\per\angstrom}$ in the $-X_3$ direction, the AB-stacked regions occupy $\approx 41\%$ of the total simulation domain in the $0.299263^\circ$ twisted bilayer hBN compared to the $\approx 38\%$ in $0.4219\%$ equi-biaxial heterostrained bilayer hBN. This corresponds to a $3\%$ smaller AB fraction in the heterostrained bilayer, highlighting the more pronounced area change due to spiral dislocation formation.

To predict the structural relaxation in the presence of the electric field, the GSFE alone is insufficient as the AB and BA domains are energetically equivalent. Area changes due to the electric fields are governed by the polarization landscape (PL) --- defined as the polarization density as a function of relative translation between the layers. Therefore, characterizing the PL is necessary to correlate the applied electric field with the spatial variation of the stacking.
\begin{figure}[t]
    \centering
\includegraphics[width=0.45\textwidth]{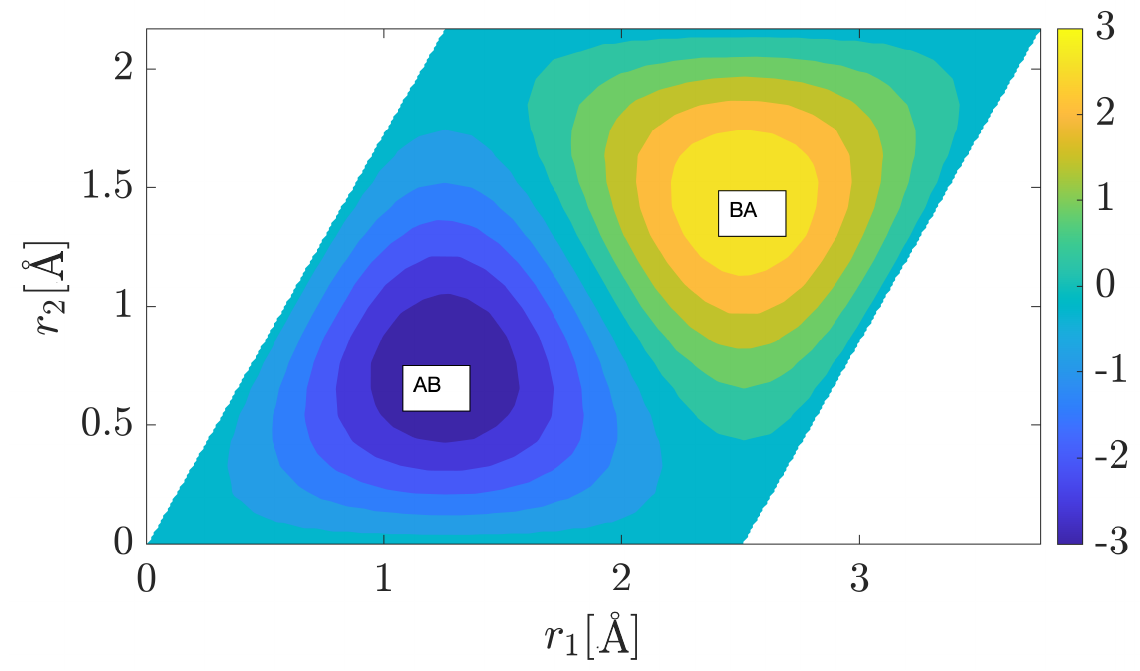}
    \caption{PL [$mC/m^2$], computed using DFT, and plotted as functions of the relative displacement between the two layers of an AA-stacked bilayer hBN.
    }
    \label{fig:pol_small}
\end{figure}
\Cref{fig:pol_small} shows the out-of-plane polarization density map of AA-stacked bilayer hBN computed using DFT. For completeness, we also show the in-plane polarization density in Section S$-4$ of the Supporting Information, which is comparable in magnitude to the out-of-plane polarization, indicating the potential for exhibiting strong in-plane ferroelectricity in small-heterodeformed bilayer hBN.\footnote{Our DFT-calculated polarization is in agreement with that of \citet{bennett2023polar}.} For comparison, the polarization densities of the AB and BA stackings relative to the AA stacking, measured using LAMMPS, are $0.264 \times 10^{-3} C/m^2$ and $-0.265 \times 10^{-3} C/m^2$ respectively. Conversely, the corresponding DFT measurements from \Cref{fig:pol_small} are $-3 \times 10^{-3} C/m^2$ and $3 \times 10^{-3} C/m^2$, which are $\tilde 10$ times the LAMMPS measurements with an opposite trend. We attribute this discrepancy to the assumption of localized charges in atomistic calculations of polarization, as also illustrated by \citet{riniker2018fixed}. Moreover, in our atomistic simulations, an electric field as high as $\SI{1}{\volt\per\angstrom}$ is required to observe noticeable aerial changes in AB/BA sackings. On the other hand, many experimental studies \cite{lv2022spatially,fan2025edge} measure similar aerial changes at lower electric fields of $0.016 V/\si{\angstrom}$. 

Such discrepancies between atomistic simulations and experiments highlight the limitations of interatomic potentials and motivate the development of DFT-informed multiscale modeling.

From \Cref{fig:pol_small}, we observe that the AA stacking has a zero out-of-plane polarization, while the AB and BA stackings are oppositely polarized with equal magnitude in the out-of-plane direction. Therefore, the equi-sized AB and BA domains in a relaxed small twisted hBN with no external electric field ensure the homostructure is unpolarized.
In the presence of an out-of-plane electric field in the $+X_3$ direction, the AB domains (positively polarized) grow and the BA domains shrink while the AA regions stay put, which is a consequence of the work done by the external electric field, first reported by \citet{yasuda2021stacking}. The evolution of the domains is mediated by the bending of the dislocation lines \cite{weston2022interfacial}. Due to the preferential growth of the AB domains, the homostructure attains a net positive polarization.

Based on the above discussion of moir\'e ferroelectricity in small heterodeformed hBN, we identify the following sufficient conditions for a bilayer configuration to demonstrate ferroelectricity:
\begin{enumerate}
    \item The heterdeformation should be small relative to a low-energy stacking up to relative translations.\footnote{We defined the energy of a bilayer's stacking as the minimum interface energy over all relative translations between its layers.} All heterodeformations introduced to this point are relative to the AA stacking.  
    \item The GSFE of the low-energy stacking has degenerate minima, i.e., the minima are energetically equivalent.
    \item The configurations corresponding to the two minima are oppositely polarized.
\end{enumerate}
The first two conditions are responsible for dislocation-mediated structural relaxation, while the last condition ensures moir\'e ferroelectricity. In a recent work, \citet{ahmed2024bicrystallography} showed that an AB-stacked bilayer graphene, twisted by $21.786789^{\circ}$, satisfies the first two conditions, and small heterodeformations relative to this configuration result in structural reconstruction mediated by interface dislocations. In the next section, we show that the PL of the $21.786789^{\circ}$ configuration satisfies condition 3, and proceed to explore ferroelectricity in large-twist hBN.

\subsection{Ferroelectricity in large-twist bilayer hBN}
\label{sec:large}
In the previous section, we identified various features of GSFE and the polarization landscape of AA-stacked bilayer hBN that lead to dislocations-mediated structural relaxation and ferroelectricity in small-heterodeformed bilayer hBN. As the heterodeformation increases, structural relaxation decreases, and interpreting the relaxed structure in terms of dislocations relative to the AA stacking breaks down as the defect cores overlap \cite{Annevelink_2020}. Interestingly, however, the authors demonstrated in an earlier work that the dislocations interpretation reemerges for small heterodeformations relative to the defect-free $21.786789^\circ$ twisted bilayer hexagonal systems under an out-of-plane compression. In particular, they showed that the interface energy of a $21.786789^\circ$ twisted bilayer graphene, minimized for local relative translations, is a local minimum with respect to the twist angle. Under small heterodeformations of this configuration, the system relaxes by nucleating interface dislocations whose Burgers vector is $\frac{1}{\sqrt{7}}$ times the Burgers vector observed in small-heterodeformed bilayer hBN configurations. In this section, we investigate whether small heterodeformations applied to the $21.786789^\circ$ twisted bilayer hBN, which itself represents a large heterodeformation relative to the AA-stacked bilayer hBN, can induce ferroelectricity.

\begin{figure}[t]
    \centering
    \subfloat[Top view]
    {
    \includegraphics[width=0.5\textwidth]{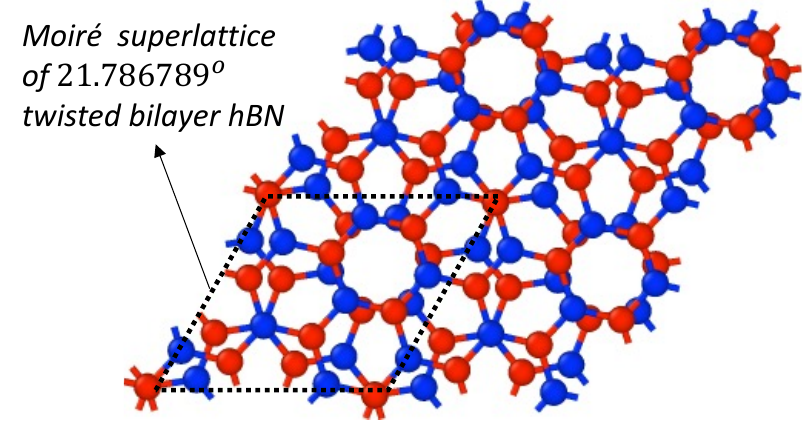}
    \label{fig:top_21_78}
    }
    \\
    \subfloat[Side view]
    {
    \includegraphics[width=0.45\textwidth]{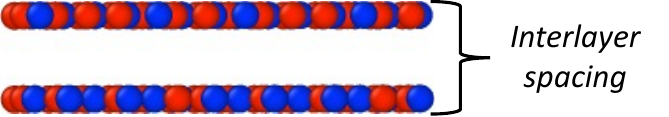}
    \label{fig:side_21_78}
    }
    \caption{Atomic configuration of a $21.787^o$ twisted bilayer hBN with red and blue colors indicating boron and nitrogen atoms, respectively. A primitive unit cell of the moir\'e superlattice, identified by the dashed lines, consists of $28$ atoms.}
        \label{fig:configuration_large_atom}
\end{figure}
\Cref{fig:configuration_large_atom} shows the atomic configuration of $21.786789^\circ$ twisted bilayer hBN.  The supercell or the coincident site lattice (CSL) of this configuration is 7 times the unit cell of hBN, and contains $28$ atoms. We will refer to this large twist bilayer hBN as the $\Sigma 7$  configuration.  
\begin{figure}[t]
    \centering
    \subfloat[]
    {
    \includegraphics[width=0.5\textwidth]{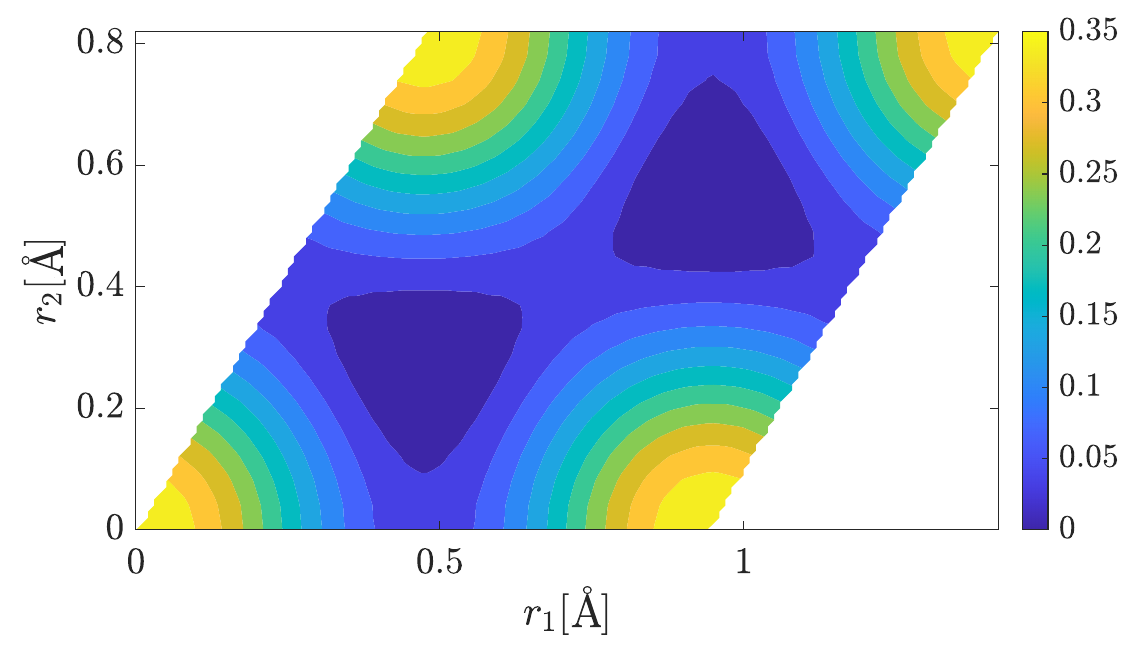}
        \label{fig:gsfe_dft_large}
    }\\
    \subfloat[]
    {
\includegraphics[width=0.5\textwidth]{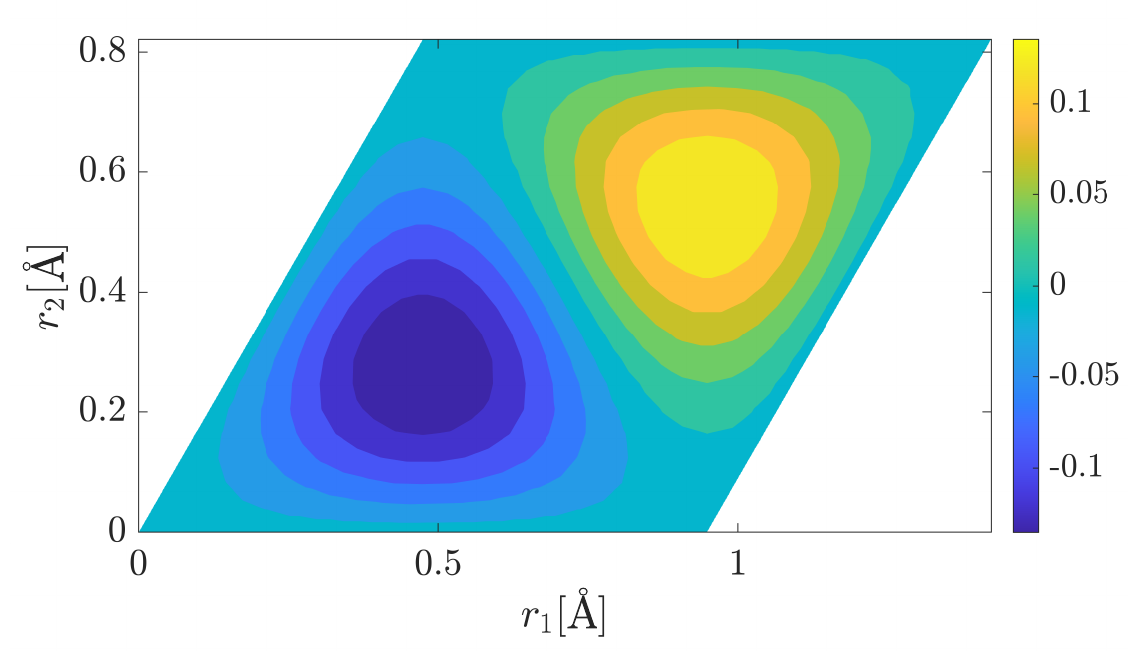}
    \label{fig:pol_large}
    } 
    \caption{\protect\subref{fig:gsfe_dft_small} GSFE [$meV\si{\angstrom}^{-2}$] and \protect\subref{fig:pol_large} PL [$mC/m^2$] density map plots of a $21.787^\circ$-twisted bilayer hBN, calculated using DFT.}
    \label{fig:gsfe_pol_large_dft}
\end{figure}
Before plotting the GSFE and the PL plots, we first infer their periodicity using the SNF bicrystallography formulated by the first and last authors in \citet{ADMAL2022}. Specifically, the GSFE and the PL plots are periodic with respect to the $\Sigma 7$ configuration's displacement shift complete lattice (DSCL), which is the coarsest lattice that contains lattices $\mathcal A$ and $\mathcal B$.  We compute the GSFE and PL of the $\Sigma 7$ bilayer hBN on the domain spanned by the DSCL basis vectors $\bm b_1 = (0.948690 \, \bm e_1)\si{\angstrom}$ and $\bm b_2=(0.474345 \,  \bm e_1 +0.821590\, \bm e_2)\si{\angstrom}$. \Cref{fig:gsfe_pol_large_dft} shows the GSFE and PL plots, computed using DFT, of the $\Sigma 7$ configuration under a $28\%$ out-of-plane compression. The two low-energy minima for the $21.786789^\circ$ twisted bilayer hBN appear only under out-of-plane compression. In the absence of compression, the generalized stacking fault energy (GSFE) surface (Supporting Information S$-5$, Figure 2a) shows a weak energy barrier between different stackings and no neighboring degenerate minima, which are required for the formation of partial dislocations \cite{ahmed2024bicrystallography}.

\begin{figure}[t]
    \centering
    \includegraphics[width=0.5\textwidth]{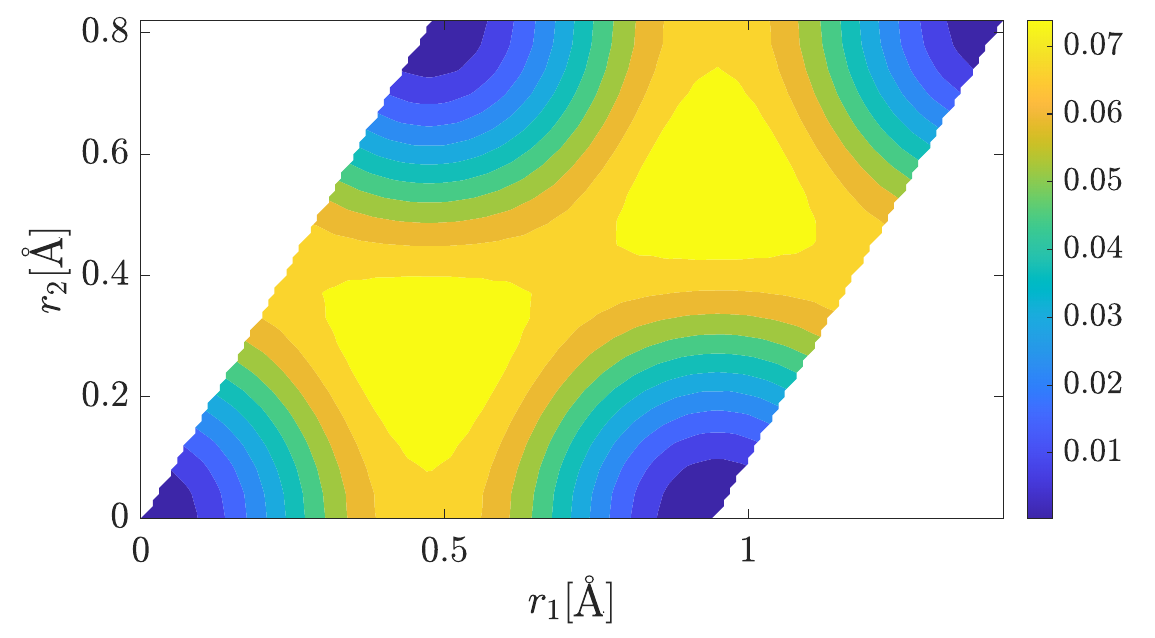}
        \caption{The GSFE [$meV\si{\angstrom}^{-2}$] of a $21.787^o$-twisted bilayer hBN computed using LAMMPS differs significantly from its DFT-counterpart in \Cref{fig:gsfe_dft_large}.}
        \label{fig:gsfe_large_atom}
\end{figure}

The GSFE in \Cref{fig:gsfe_dft_large} shows two degenerate minima similar to that of the $0^\circ$ bilayer hBN in \Cref{fig:gsfe_small_dft_atom}, suggesting small heterodeformations relative to the $\Sigma 7$ configuration will result in interface dislocations. The size of the Burgers vectors of such dislocations is given by the distance between the minima, which is $\approx 0.55 \si{\angstrom}$. Moreover, analogous to the small twist case, since each minimum in \Cref{fig:gsfe_dft_large} is surrounded by three minima, we expect the interface dislocations to form a triangular network. While in an earlier work, we demonstrated structural relaxation in heterodeformed $\Sigma 7$ bilayer graphene using atomistic simulations, we are unable to repeat such a calculation for bilayer hBN in this work since the ILP potential is not applicable for large out-of-plane compressions. \Cref{fig:gsfe_large_atom} reflects the failure of the ILP potential --- not only is the magnitude of the LAMMPS-calculated GSFE inconsistent with that in \Cref{fig:gsfe_dft_large}, but also the distribution of the extrema does not agree. 

The PL plot in \Cref{fig:pol_large} shows that the $\Sigma 7$ stacking has a zero out-of-plane polarization density, while the stackings corresponding to the degenerate minima have opposite polarization densities of magnitude $1.35\times 10^{-4} C/m^2$.  Therefore, all three sufficient conditions for ferroelectricity, outlined at the end of \Cref{sec:small}, are met by large-heterodeformed bilayer hBN vicinal to the $\Sigma 7$ configuration.

The potential of ferroelectricity under large heterodeformations identified in this section and the absence of reliable interatomic potentials motivate us to develop a DFT-informed multiscale framework that is computationally efficient compared to atomistic simulations and capable of capturing large-twist bilayer hBN physics. In the next section, we present the bicrystallography-informed and frame-invariant multiscale model for the prediction of ferroelectricity in arbitrarily heterodeformed bilayer hBN.
\section{A multiscale model for ferroelectricity in heterodeformed hBN}
\label{sec:cont_fram}
In this section, we present a bicrystallography-informed frame-invariant multiscale (BFIM) model for predicting ferroelectricity in arbitrarily heterodeformed bilayer hBN. The goal of this model is to predict the structural response and the polarization density field when a bilayer hBN is subjected to an electric field. In this work, we extend the GFK model of \citet{ahmed2024bicrystallography}, which was used to predict structural relaxation, to include polarization.

In the BFIM model, a bilayer is described as two continuum sheets. It consists of  a) a defect- and strain-free configuration called the \emph{natural configuration}, which is used to construct the energy of the system, b) a \emph{reference configuration}, with respect to which displacements are measured, and a c) \emph{deformed configuration} which represents the deformed bilayer. If a heterodeformation is vicinal to the AA stacking, then the AA stacking is chosen as the natural configuration. On the other hand, if a heterodeformation is vicinal to the $\Sigma 7$ configuration, then the $\Sigma 7$  configuration is chosen as the natural configuration. A small heterodeformation is introduced by uniformly deforming one of the layers, resulting in the reference configuration. The model is designed to predict the displacements from the reference configuration  to the deformed configuration arising due to atomic reconstruction. 

\subsection{Kinematics}
In the reference configuration, the two layers of the bilayer are represented by subsets $\Omega^{\rm ref}_{\rm t}$ and $\Omega^{\rm ref}_{\rm b}$ of the 2D Euclidean point space $\mathbb R^2$. An arbitrary material point in the bilayer is denoted by $\bm X_{\rm t}$ or $\bm X_{\rm b}$, depending on whether the point belongs to the top or the bottom layer. In this study, we ignore out-of-plane displacement since ferroelectricity is primarily governed by the in-plane reconstruction \cite{weston2022interfacial}. $\bm \phi_\alpha:\Omega^{\rm ref}_\alpha \times [0,\infty] \to \mathbb R^2$ ($\alpha=\rm t,\rm b$)\footnote{\rm t and \rm b stand for top and bottom layers, respectively} denote time-dependent deformation maps associated with the atomic reconstructions of the respective lattices. The images of the maps $\bm \phi_{\rm t}$ and $\bm \phi_{\rm b}$ constitute the deformed configuration. At $t=0$, we assume $\bm \phi_\alpha(\bm X_\alpha,0)=\bm X_\alpha$. Therefore, $\bm \phi_\alpha(\bm X_\alpha,t)-\bm X_\alpha$ describes the displacement field associated with atomic reconstruction. 

The lattice strain in the deformed configuration is measured relative to the strain-free natural configuration. If $\bm \kappa_\alpha$ denotes the mapping between the reference and the natural configuration, and $\bm \eta_\alpha$ denotes the mapping from the natural configuration to the deformed configuration, we have
\begin{equation}
    \bm \phi_\alpha = \bm \eta_\alpha \circ \bm \kappa_\alpha,
    \label{eqn:composition}
\end{equation}
where $\circ$ denotes function composition. The inverse of $\bm \kappa_\alpha$ is the deformation required to form the uniformly heterodeformed bilayer hBN configuration. If $\bm F_\alpha$ denotes the gradient of $\bm \phi_\alpha$, \cref{eqn:composition} implies
\begin{equation}
   \bm F_\alpha = \bm H_\alpha \bm K_\alpha, \text{ where } \bm H_\alpha:=\nabla \bm \eta_\alpha \text{ and, } \bm K_\alpha:=\nabla \bm \kappa_\alpha.
   \label{eqn:FeFp}
\end{equation}

To construct the system's energy, we note that the total energy includes the elastic energy due to lattice strains, the interfacial van der Waal's (vdW) energy, and the work done by the external electric field through its interaction with the polarization. We build each of these energy components using frame-invariant measures. A frame-invariant kinematic measure of the lattice strain is the Lagrangian strain $\bm E_\alpha:= (\bm H_\alpha^{\rm T} \bm H_\alpha-\bm I)/2$. From \cref{eqn:FeFp}, $\bm E_\alpha$ can be written as
\begin{equation}
    \bm E_\alpha = (\bm K_\alpha^{-\rm T} \bm F^{\rm T}\bm F \bm K_\alpha^{-1}-\bm I)/2.
\end{equation}
The interfacial vdW energy and the work done by the external electric field are associated with the interface. Specifically, these energies depend on the stacking, which varies spatially and depends on the relative shift between layers. Therefore, the vdW energy and polarization energy density at a point, $\bm x$, in the deformed configuration are described as functions of the relative translation vector
\begin{equation} 
\bm r(\bm x,t)= \Kt \Xt-\Kb \Xb, \text{ where }
    \bm X_\alpha:= \bm \phi_\alpha^{-1}(\bm x,t).
    \label{eqn:r}
\end{equation}
Note that $\bm r$ is frame-invariant by design. In the next section, we will construct the constitutive law (energy and mobility) in terms of the kinematic measures $\bm C_\alpha$ and $\bm r$.

\subsection{Constitutive law}
\label{sec:constitutive}
Expressing the total energy as  $\mathcal E=\mathcal E_{\rm el}+\mathcal E_{\rm vdW} - \mathcal E_{\rm pol}$, we will now construct the elastic energy $\mathcal E_{\rm{el}}$, van der Waals energy $\mathcal E_{\rm{vdW}}$, and the work $\mathcal E_{\rm pol}$ done by the electric field as functionals of the unknown fields $\bm \phi_{\rm t}$ and $\bm \phi_{\rm b}$. The elastic energy is expressed in a Saint Venant--Kirchhoff form as 
\begin{equation}
    \mathcal E_{\rm el}[\phit, \phib] = \sum_{\alpha=\mathrm t, \mathrm b}
    \int_{\Omega_\alpha^{\rm n}} 
    e_{\rm el}(\bm E_\alpha;\alpha) \, d\bm Y_\alpha, 
    \label{eqn:elasticEnergy}
\end{equation}
where 
\begin{equation}
    e_{\rm el}(\bm E_\alpha) = \frac{1}{2} \mathbb C \bm E_\alpha \cdot \bm E_\alpha= \lambda (\mathrm{tr}\, \bm E_\alpha)^2 + 2\mu \bm E_\alpha \cdot \bm E_\alpha,
\end{equation}
is the elastic energy density of the $\alpha$-th layer, and $\mathbb C$ is the fourth-order isotropic elasticity tensor with lam\'e constants\cite{jung2015origin} $\lambda=\SI{3.5}{\eV\per\angstrom\squared}$ and $\mu=\SI{7.8}{\eV\per\angstrom\squared}$. Note that the integral in \cref{eqn:elasticEnergy} is over the natural configuration with $d\bm Y_\alpha$ representing a unit volume in the natural configuration $\Omega_\alpha^{\rm n}$. The interfacial vdW energy originates from the interaction between the layers in the region $\Omega_{\rm t} \cap \Omega_{\rm b}$. The expression of vdW energy is constructed as 
\begin{equation}
\mathcal E_{\rm vdW}[\phit, \phib] =  
\frac{1}{2}
\sum_{\alpha=\rm t, \rm b}
\int_{\Omega_{\rm t} \cap \Omega_{\rm b}} 
(\det \bm H_\alpha)^{-1} e_{\rm vdW}(\bm r(\bm x_\alpha)) \, d\bm x_\alpha,
\label{eqn:interEnergy}
\end{equation} where, $e_{\rm vdW}$ is the interfacial energy density (per unit volume in the natural configuration). Note that the factor $(\det \bm H_\alpha)^{-1}$ is necessary because the integration is over the deformed configuration as opposed to the natural configuration. The vdW energy density is the GSFE per unit area of the natural configuration. Therefore, it is expressed in a form that reflects the symmetry of the GSFE as 
\begin{equation}
    e_{\rm vdW}(\bm r) = \pm 2 v_{\rm g}\sum_{p=1}^{3}\cos \left (2 \pi 
        \bm{\mathcal d}^p \cdot \bm r
    \right)+c_{\rm g},
    \label{eq:vdW_en}
\end{equation}
where $v_{\rm g}$ is the strength of the GSFE, $\bm {\mathcal d^1}$ and $\bm {\mathcal d^2}$ are reciprocal to the two DSCL vectors that span the domain of the GSFE, and $\bm {\mathcal d}^3=-(\bm {\mathcal d}^1+\bm {\mathcal d}^2)$. 

The parameters $v_{\rm g}$ and $c_{\rm g}$ are obtained by comparing \cref{eq:vdW_en} with the atomistic GSFE of the AA stacking for small twist bilayer hBN, and the DFT GSFE for $28\%$ out-of-plane compressed large-heterodeformed bilayer hBN.

The work done by the electric field originates from its interaction with the polarized interface $\Omega_{\rm t} \cap \Omega_{\rm b}$. Therefore, the functional $\mathcal E_{\rm pol}$ is constructed as 
\begin{equation}
\mathcal E_{\rm pol}[\phit, \phib] =  
\frac{1}{2}
\sum_{\alpha=\rm t, \rm b}
\int_{\Omega_{\rm t} \cap \Omega_{\rm b}} 
(\det \bm H_\alpha)^{-1} e_{\rm pol}(\bm r(\bm x_\alpha)) \, d\bm x_\alpha,
\label{eqn:polEnergy}
\end{equation} 
where $e_{\rm pol}$ is the work done by an external electric field $\bm E$. It is of the form 
\begin{equation}
    e_{\rm pol}(\bm r) = \chi^{-1} \bm E.\bm P,
    \label{eq:pol_en}
\end{equation}
where $\bm P$ denotes the spatially varying dipole moment that depends on the local stacking and is obtained from the PL plots. The out-of-plane component of the polarization density field can be represented as 
\begin{equation}
    \bm P = \pm 2 \mathfrak{d} v_f\sum_{p=1}^{3}\sin \left (2 \pi 
        \bm{\mathcal d}^p \cdot \bm r
    \right)+c_f,
    \label{eqn:polarizationDensity}
\end{equation}
where $v_f$ is the maximum polarization, and $\mathfrak{d}$ is the interlayer spacing between the layers. The parameter $v_f$ is obtained by comparing \cref{eqn:polarizationDensity} with the atomistic PL of the AA stacking for small twist bilayer hBN, and the DFT PL for $28\%$ out-of-plane compressed large-heterodeformed bilayer hBN.\footnote{Incorporation of $28\%$ out-of-plane compressed GSFE and PL is essential to simulate the accurate relaxed structure in the presence and absence of applied electric field.} The conversion factor $\chi$ in \cref{eq:pol_en} is obtained by comparing the areal changes in AB and BA domains, predicted by the BFIM model, to those reported by \citet{lv2022spatially} in a $0.3^\circ$ twisted bilayer hBN. \citet{weston2022interfacial} used a similar approach to fit their model studying structural relaxation in twisted bilayer molybdenum disulfide under an applied electric field.

To simulate structural relaxation, we minimize the total energy using the gradient flow
\begin{equation}
    m \dot{\bm \phi}_\alpha = -\updelta_{\bm \phi_\alpha} \mathcal E, \quad (\alpha = \rm t, \rm b)
    \label{eqn:gradientFlow}
\end{equation}
where $m\equiv 1$ is the mobility associated with $\bm \phi_\alpha$s, and $\updelta_{\bm \phi_\alpha}$ denotes the variation with respect to $\bm \phi_\alpha$. 
\subsection{Derivation of governing equations}
In this section, we derive the governing equations for the dynamics of structural relaxation in heterodeformed bilayer hBN. Computing the variational derivatives of \cref{eqn:elasticEnergy,eqn:interEnergy,eqn:polEnergy}, and substituting them into \cref{eqn:gradientFlow}, we obtain
\begin{subequations}
\label{eq:final_eqn}
\begin{align}
    m \dot{\bm \phi}_{\rm t} &= \divrt(\bm P_{\rm t}) + \Hb^{-\rm T} \nabla e_{\rm vdW}(\bm r_{\rm t})- \Hb^{-\rm T} \nabla e_{\rm pol}(\bm r_{\rm t}), \\
    m \dot{\bm \phi}_{\rm b} &= \divrb(\bm P_{\rm b}) -  \Ht^{-\rm T} \nabla e_{\rm vdW}(\bm r_{\rm b}) + \Ht^{-\rm T} \nabla e_{\rm pol}(\bm r_{\rm b}).
\end{align}
\end{subequations}
where, $\bm P_\alpha:= \bm H_\alpha \bm \nabla \bm e_{\rm el} \bm K_\alpha^{-\rm T}$, is the 2D analog of the elastic Piola--Kirchhoff stress, which measures force in $\Omega_\alpha$ measured per unit length in $\Omega_\alpha^{\rm ref}$. Furthermore, $\rt$ and $\rb$ are given by \footnote{See Section 4.3 in \citep{ahmed2024bicrystallography} for the fully general expressions for $\bm r_{\rm t}$ and $\bm r_{\rm b}$, arguments for their approximations in \eqref{eqn:r_approx}, and boundary conditions relevant for finite systems.} , 
\begin{subequations}
\begin{align}
    \rt &\approx (\Kt -\Kb) \Xt -  \Hb^{-1} (\phit(\Xt)-\phib(\Xt)),\label{eqn:rt_approx}\\
    \rb &\approx (\Kt -\Kb) \Xb -  \Ht^{-1} (\phit(\Xb)-\phib(\Xb)). \label{eqn:rb_approx}
\end{align}
\label{eqn:r_approx}
\end{subequations}

Details about the numerical implementation of \eqref{eq:final_eqn} are given in Section S$-2$ of the Supporting Information. Although the BFIM model is applicable for both finite and periodic boundary conditions, we limit its implementation to PBCs in this paper because the goal is to compare bulk scale experimental observations of \citet{lv2022spatially}. 

In the next section, we will use the BFIM model to predict structural reconstruction with and without an applied electric field for any arbitrarily heterodeformed bilayer hBN.   
\section{Results and Discussion}
\label{sec:results}
This section shows simulation results of the BFIM model demonstrating how structural relaxation occurs under an external electric field in various small and large heterodeformed bilayer hBNs. Additionally, we compare the structural relaxations predicted by the BFIM model to those observed in experiments under out-of-plane electric fields
\begin{figure*}
    \centering
    \subfloat[]
    {
    \includegraphics[width=0.5\textwidth]{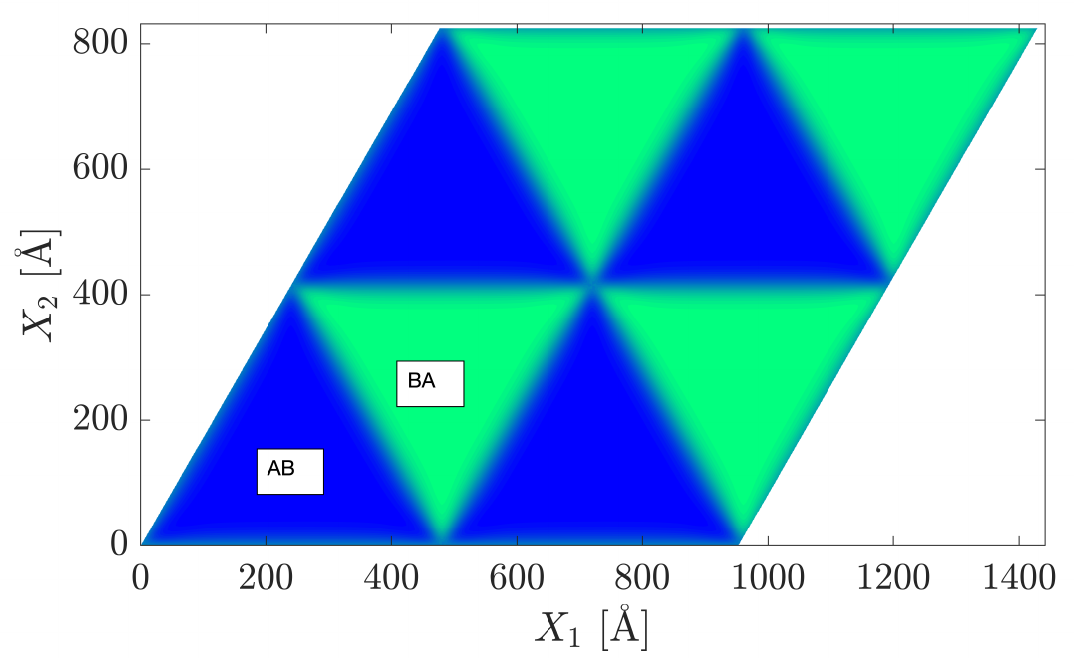}
    \label{fig:relax_029_cont}
    }
    \subfloat[]
    {
\includegraphics[width=0.5\textwidth]{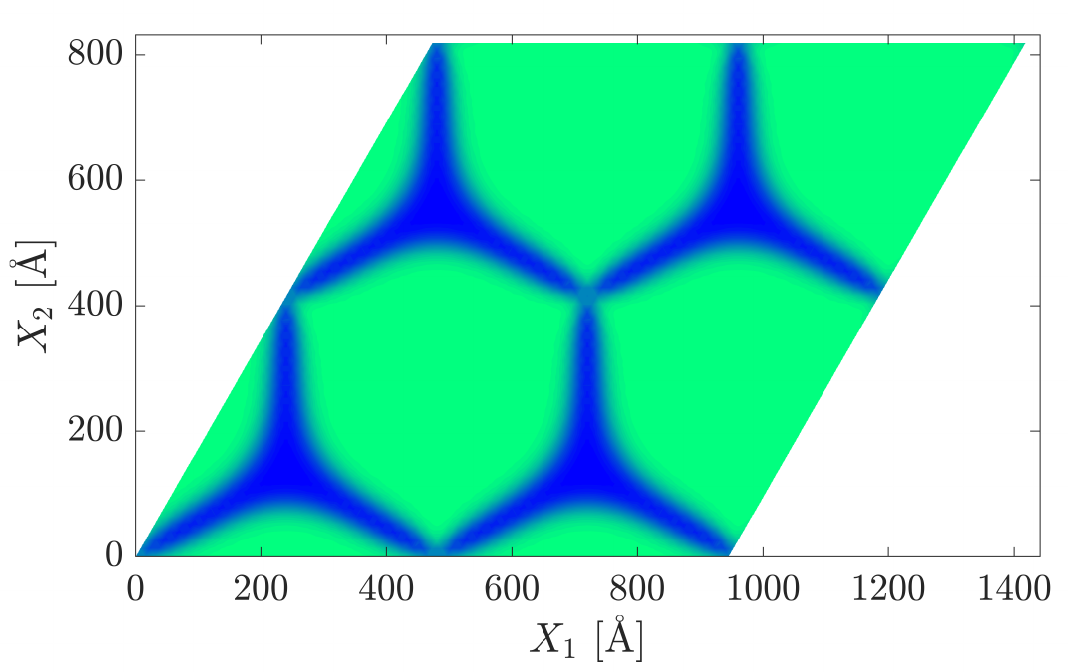}
    \label{fig:pos_5E_cont}
    }
    \\
    \subfloat[]
    {
\includegraphics[width=0.5\textwidth]{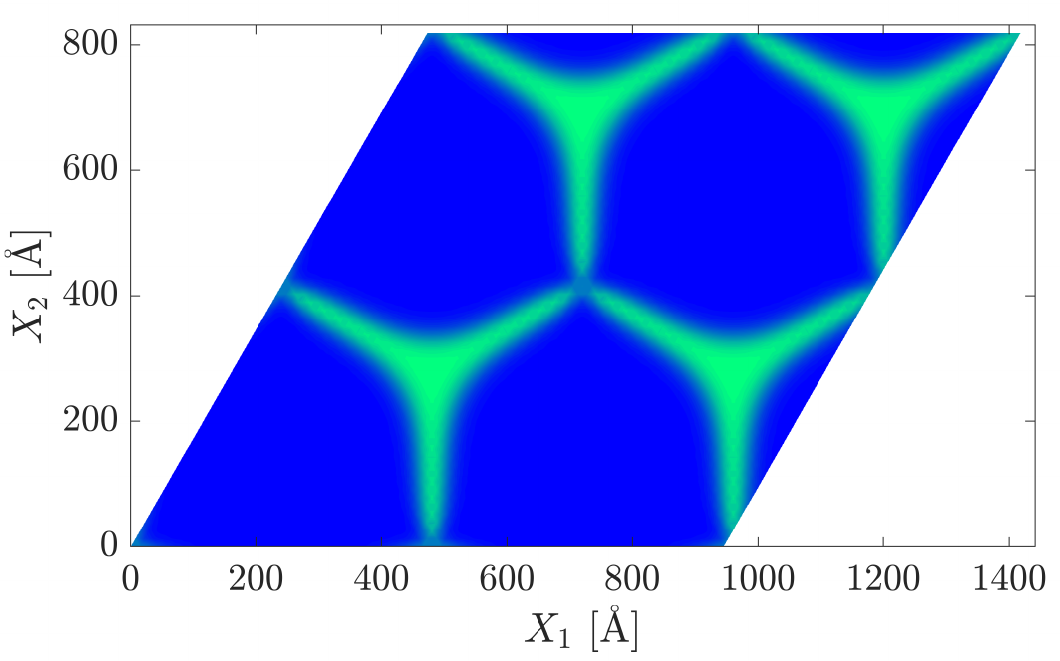}
        \label{fig:neg_5E_cont}
    }
    \subfloat[]
    {
\includegraphics[width=0.45\textwidth]{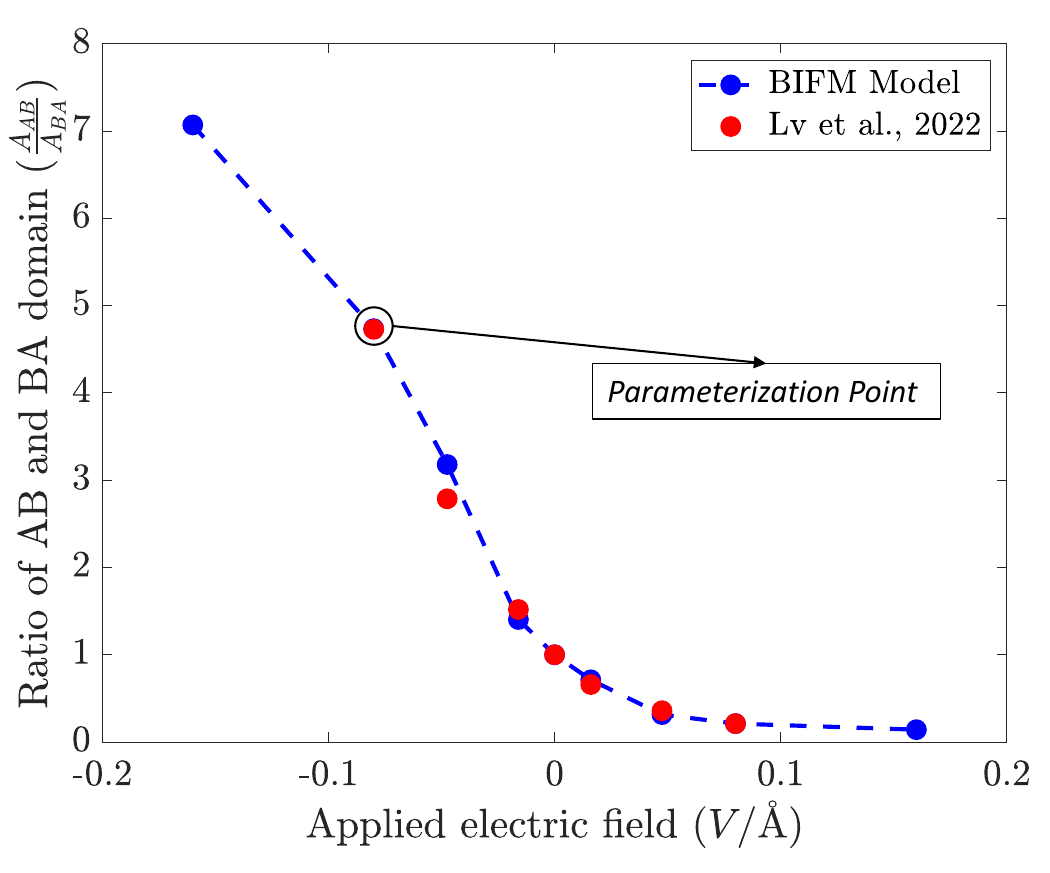}
    \label{fig:cont_comp_twist}    
    } 
    \caption{BFIM model prediction of ferroelectric domain formation in a $0.299^o$ twisted bilayer hBN. The blue and green regions in the polarization density contour plots depict the AB and BA domains.  \protect\subref{fig:relax_029_cont}, \protect\subref{fig:pos_5E_cont}, and \protect\subref{fig:neg_5E_cont} correspond to $E=0$, $+\SI{0.08}{\volt\per\angstrom}$, $-\SI{0.08}{\volt\per\angstrom}$, respectively. The latter two closely resemble the atomistic simulation plots \cref{fig:ferro_atom_twist}. \protect\subref{fig:cont_comp_twist} compares the areal change ratio, $\frac{A_{AB}}{A_{BA}}$, illustrated in \protect\subref{fig:relax_029_cont}, with that reported by \citet{lv2022spatially} for various electric fields.}
    \label{fig:cont_atom_comp}
\end{figure*} 
We begin by first fitting $\chi$, the only unknown parameter of the BFIM model, by simulating the structural relaxation of a $0.299263^{\circ}$ twisted bilayer hBN at $E=-0.08V/\si{\angstrom}$. \Cref{fig:cont_atom_comp} shows continuum simulation results of structural relaxation at $E=0$, $E=+\SI{0.08}{\volt\per\angstrom}$, and $E=-\SI{0.08}{\volt\per \angstrom}$. As expected, the model predicts the triangular network of dislocations for $E=0$, and domains that shrink (grow) under $E=+\SI{0.08}{\volt\per\angstrom}$ grow (shrink) when the sign of the electric field switches. The parameter $\chi$ is fit using a mid-point search strategy such that the areal ratio, $A_{AB}/A_{BA}$ at $E=-\SI{0.08}{\volt\per\angstrom}$, matches with the experimental measurement of \citet{lv2022spatially}. The fitting yields the $\chi^{-1}=11$. Next, we validate the parametrized model by comparing (see \cref{fig:cont_comp_twist}) its predictions of $A_{AB}/A_{BA}$ for various applied electric fields with experimental measurements of \citet{lv2022spatially}.
\Cref{fig:cont_comp_twist} shows good agreement between BFIM model predictions and experiment while demonstrating the non-linear dependence of $A_{AB}/A_{BA}$ on the applied field. In contrast to \cref{fig:ferro_atom_twist}, the AB(BA) domain increases(decreases) in area under a negative electric field, which is consistent with the PL plot of AA stacked bilayer hBN. This discrepancy in atomistic simulations is a consequence of the incorrect sign and magnitude of the LAMMPS-calculated polarization densities of AB and BA stackings, which required a significantly strong electric field to deform the interface dislocations compared to electric fields as small as $0.0475V/\si{\angstrom}$ used in experiments.
\begin{figure} 
    \centering 
\includegraphics[width=0.5\textwidth]{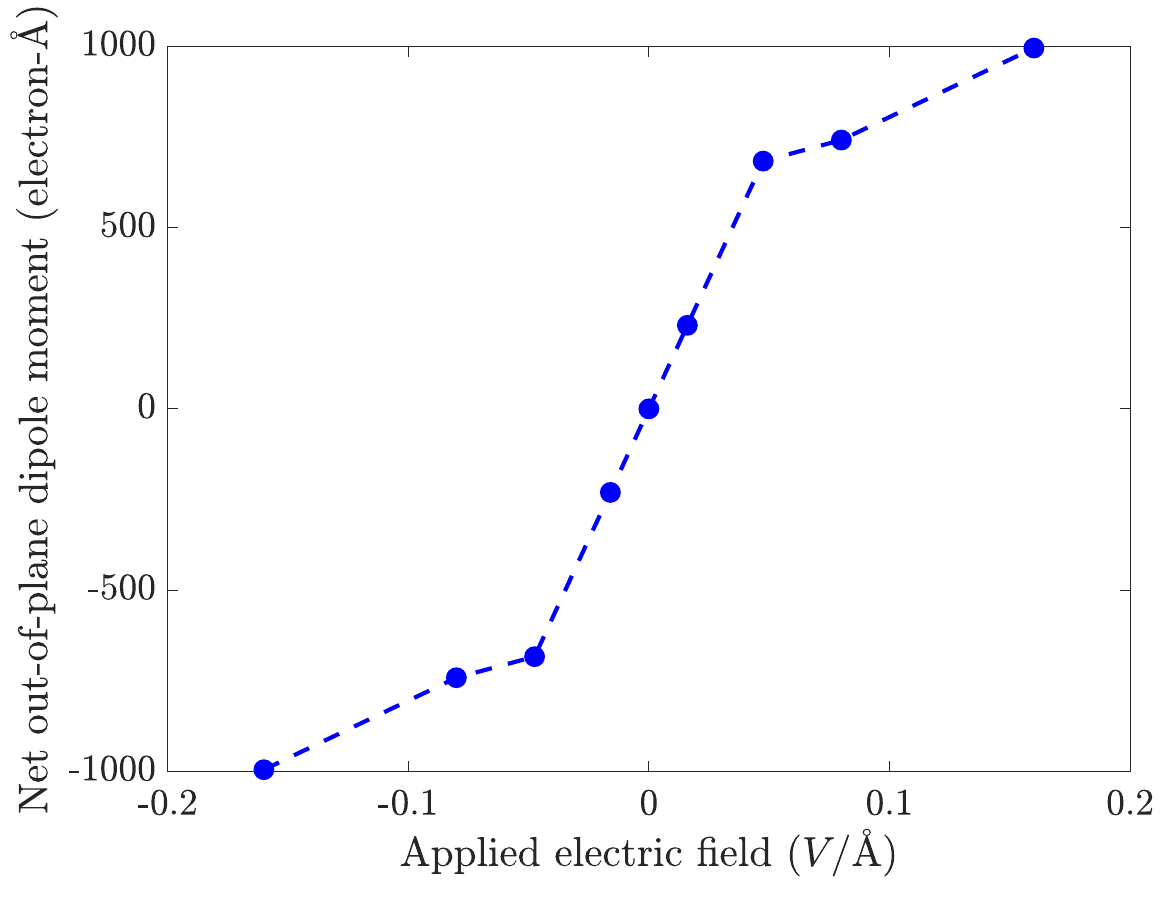} 
    \caption{Variation of net out-of-plane dipole moment with applied electric field in a $0.299^o$ twisted bilayer hBN, computed using the BFIM model. 
    } 
    \label{fig:con_dip_moment} 
\end{figure}
To further illustrate this field-dependent response, \cref{fig:con_dip_moment} shows the variation of the net out-of-plane dipole moment with electric field in a $0.299^\circ$ twisted bilayer hBN, computed using the BFIM model. From the figure, we note that the dipole moment increases nonlinearly with electric field with a rapid increase at lower electric field magnitudes. Such trends are consistent with those reported by \citet{bennett2022electrically} in twisted bilayer molybdenum disulfide.   
\begin{figure*}
    \centering
    \subfloat[]
    {
\includegraphics[width=0.5\textwidth]{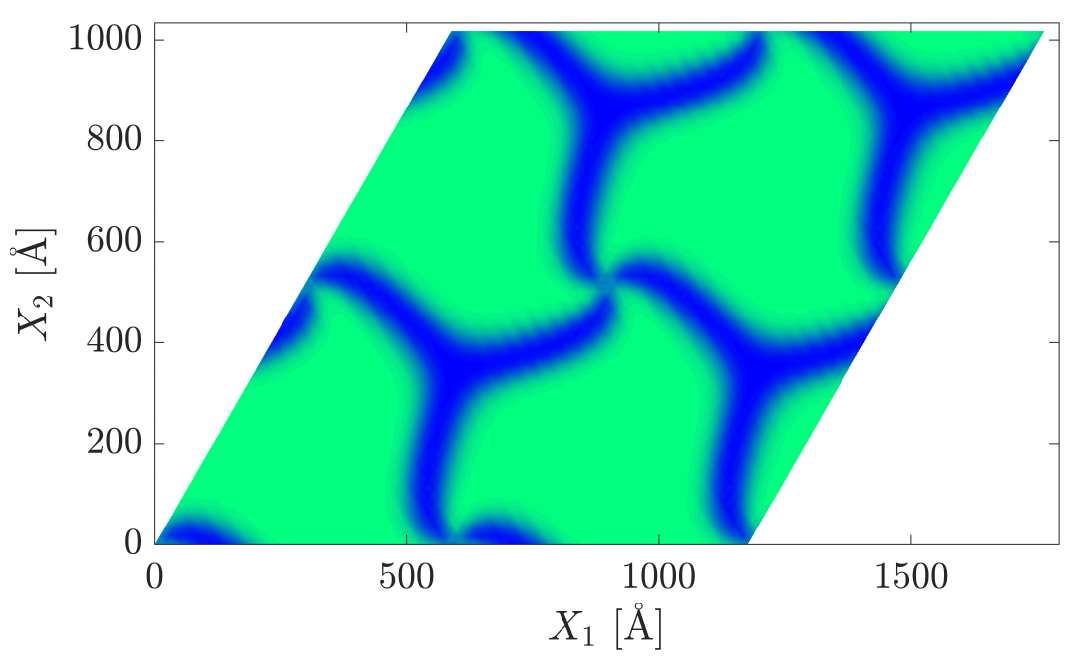}
    \label{fig:str_pos_cont}
    }
    \subfloat[]
    {
\includegraphics[width=0.5\textwidth]{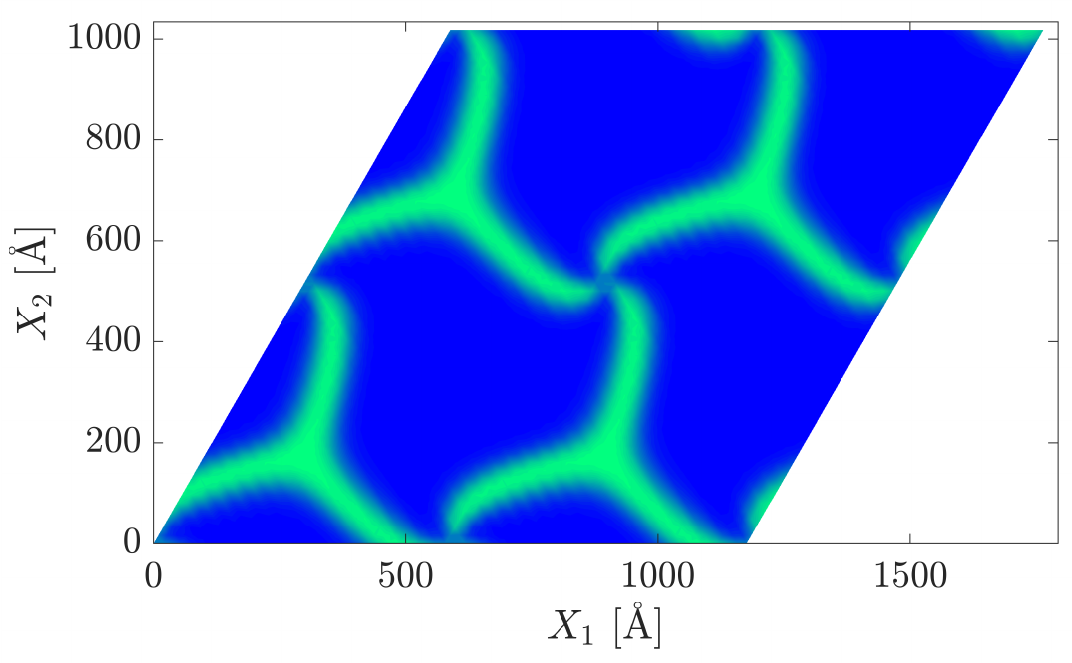}
        \label{fig:str_neg_cont}
    } 
    \caption{BFIM model prediction of ferroelectric domain formation in $0.422\%$ equi-biaxial heteorostrained bilayer hBN under applied electric field of \protect\subref{fig:str_pos_cont} $+0.08 V/\si{\angstrom}$ and, \protect\subref{fig:str_neg_cont} $-0.08 V/\si{\angstrom}$.}
    \label{fig:str_results_cont}
\end{figure*}

Moving on to small heterostrains, \cref{fig:str_results_cont} shows ferroelectric domain formation in a $0.4219\%$ equi-biaxial heterostrained hBN under alternating positive and negative electric fields. From the plots, it is clear that the BFIM model accurately predicts the formation of spiral dislocations and their response to electric fields, as noted in \cref{fig:ferro_atom_str}.

\begin{figure*}
    \centering
    \subfloat[]
    {
\includegraphics[width=0.5\textwidth]{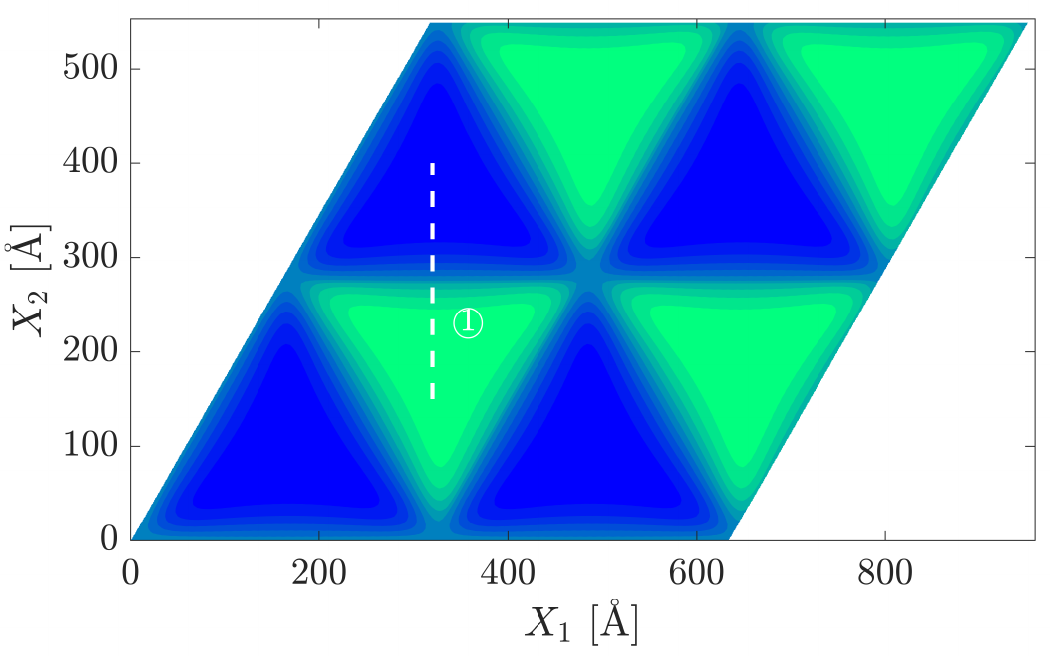}
    \label{fig:large_relax_cont}
    }
    \subfloat[]
    {
\includegraphics[width=0.5\textwidth]{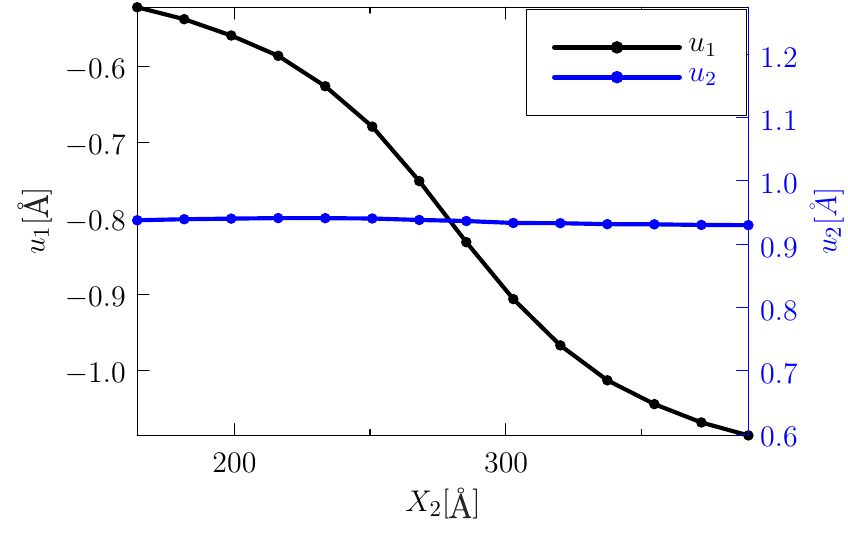}
        \label{fig:large_relax_Burgers}
    } 
    \caption{\protect\subref{fig:large_relax_cont} Relaxed configuration of $21.957^o$ twisted bilayer hBN at $28\%$ out-of-plane compression, computed using BFIM model, showing an array of dislocations whose Burgers vector is shown in \protect\subref{fig:large_relax_Burgers} through the displacement components along scanning direction \textcircled{1} shown in \protect\subref{fig:large_relax_cont}}
\label{fig:relax_twist_large_cont}
\end{figure*}

Finally, we will use the BFIM model to demonstrate ferroelectric domain formation in large heterodeformed bilayer hBN. Recall from \Cref{sec:large}, we expect that any small heterodeformation relative to the $\Sigma 7$ twisted bilayer hBN configuration will result in interface dislocations-mediated structural reconstruction and ferroelectricity. Therefore, using SNF bicrystallography, we choose the $21.786789^\circ+0.170076^\circ$ twisted bilayer hBN as a case study. 
\begin{align}
    \bm b_1 = (319.600 \, \bm e_1)\si{\angstrom}, 
    &\quad 
    \bm b_2=(159.800 \,  \bm e_1 + 276.781\, \bm e_2)\si{\angstrom}.
\end{align}
\Cref{fig:large_relax_cont} shows the relaxed $21.956865^\circ$ twisted bilayer hBN at $28\%$ out-of-plane compression, consisting of a triangular network of interface dislocations. Although the dislocations appear similar to those in a small-twist bilayer hBN, the Burgers vector is markedly different. \Cref{fig:large_relax_Burgers} shows the displacement components along the scanning direction \textcircled{1}. The displacement component normal to the line direction is negligible, while the component along the line direction is $\approx0.55 \si{\angstrom}$, which implies the dislocations have a screw character with Burgers vector magnitude $\approx0.55 \si{\angstrom}$. Recall that this magnitude is equal to the distance between the minima of the $\Sigma 7$ GSFE plotted in \cref{fig:gsfe_pol_large_dft}. To estimate the minimum energy required to initiate dislocation motion, we compute the Peierls stress for interface dislocations (Supporting Information S$-6$), which is negligible, indicating that ferroelectric switching via dislocation motion occurs nearly instantaneously.

\begin{figure*}
    \centering
    \subfloat[]
    {
\includegraphics[width=0.5\textwidth]{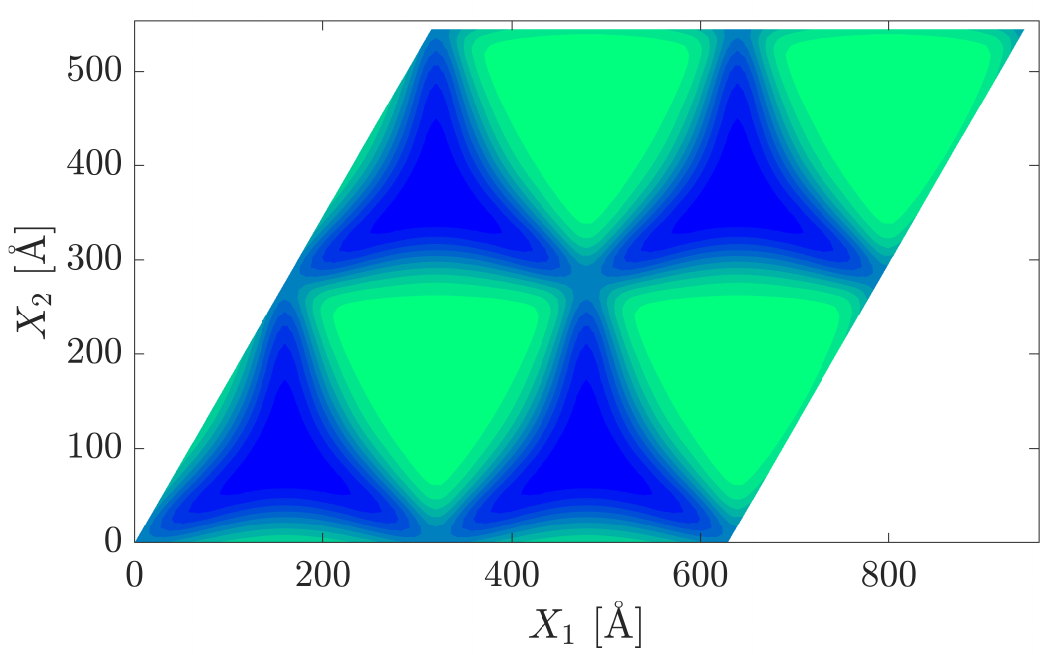}
    \label{fig:large_pos}
    }
    \subfloat[]
    {
\includegraphics[width=0.5\textwidth]{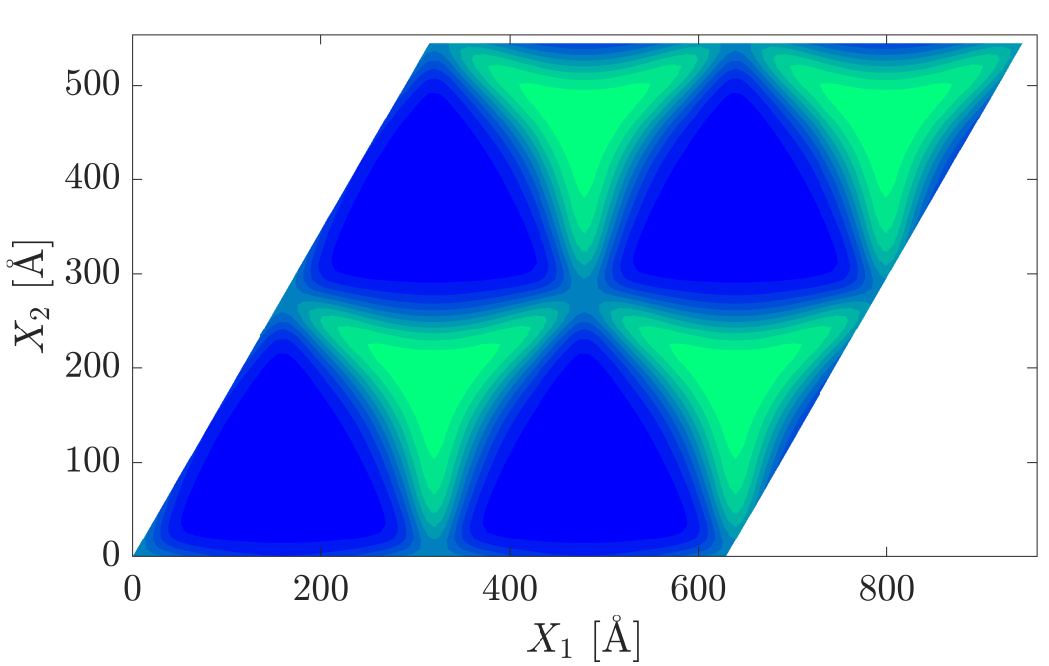}
        \label{fig:large_neg}
    } 
    \caption{BFIM model prediction of ferroelectric domain formation in $21.957^o$ twisted bilayer hBN at $28\%$ out-of-plane compression under applied electric field of \protect\subref{fig:large_pos} $+0.16 V/\si{\angstrom}$ and, \protect\subref{fig:large_neg} $-0.16 V/\si{\angstrom}$.}
    \label{fig:large_ferro_domain}
\end{figure*}

From the polarization landscape in \cref{fig:pol_large}, we know that the domains in \Cref{fig:large_relax_Burgers} are oppositely polarized. Therefore, as in the small-twist case, \Cref{fig:large_ferro_domain} shows that the domains grow/shrink under an external electric field. However, the areal change ratio, $\frac{A_{AB}}{A_{BA}}$ is $\approx 4$ times smaller than the one observed in $0.299^\circ$ twisted bilayer hBN under electric field of $-0.16V/\si{\angstrom}$. This reduction in areal change is a consequence of the reduced PL strength of large-twist bilayer hBN. 

\section{Summary and conclusions}
\label{sec:conclusions}
It is well known that 2D bilayers undergo structural reconstruction when a small relative twist is applied between their layers, leading to alternating domains of low-energy stackings (AB and BA) separated by interface dislocations or strain solitons. In the case of a small-twist hBN bilayer, its two energetically equivalent AB and BA stackings are oppositely polarized. This feature results in ferroelectricity in small-twist bilayer hBN, meaning the system becomes polarized in response to an electric field, with one domain expanding or contracting at the expense of the other, depending on the direction of the electric field.

In this paper, we studied ferroelectricity in hBN beyond the small twist case by extending previous studies to arbitrary heterodeformations. To examine the small heterostrain case, we used atomistic simulations to demonstrate ferroelectricity under small-biaxial heterostrain. In particular, we demonstrated that in this case, structural relaxation also leads to alternating AB and BA domains. However, unlike the small-twist case, they are separated by swirling interface dislocations. Alternating the direction of the electric field drives the expansion and contraction of the domains, leading to ferroelectricity.  

The large-heterodeformation study was inspired by our earlier work \cite{ahmed2024bicrystallography}, which recognized that structural reconstruction occurs not only for small heterodeformations relative to the AA stacking but also for those relative to a specific twist angle of $21.786789^\circ$, which we referred to as the $\Sigma 7$ configuration. Under such large heterodeformations, we demonstrated that structural reconstruction occurs through the formation of alternate low-energy stackings, separated by interface dislocations whose Burgers vector is significantly smaller than that observed in the small-twist case. This observation led us to the question  --- \emph{does bilayer hBN demonstrate ferroelectricity under large heterodeformations in the vicinity of the $\Sigma 7$ configuration?}

The absence of a reliable interatomic potential for large heterodeformations prompted us to leapfrog the atomic scale and develop a DFT-informed continuum model --- the bicrystallography-informed frame-invariant multiscale (BFIM) model --- which applies to any heterodeformation.  Thereafter, we systematically addressed the above question by first mapping the generalized stacking fault energy and the polarization landscape of the defect-free $\Sigma 7$ configuration using DFT calculations as functions of relative translation between the layers of the $\Sigma 7$ configuration. Analogous to the small-twist case, the maps demonstrated the presence of oppositely polarized low-energy stackings, indicating that heterodeformations in the vicinity of the $\Sigma 7$ configuration will exhibit ferroelectricity. Using the above maps as input to the continuum model, we demonstrated ferroelectricity under large heterodeformations.

Before concluding, we acknowledge the limitations of the current study. First, the BFIM model does not account for out-of-plane displacement. \citet{dai2016twisted,rakib2023helical} noted that interlayer dislocations can transform from straight to helical lines, forming out-of-plane bulges at the AA junction. Moreover, this out-of-plane bulge might even be responsible for forming spiral dislocations in equi-biaxially heterostrained bilayer hBN as shown by \citet{zhang2024impact}. To incorporate the out-of-plane displacement, the constitutive law of the BFIM model should include a) a 3D GSFE \citep{zhouVanWaalsBilayer2015} (as opposed to the current 2D GSFE), wherein the third dimension corresponds to the interlayer spacing, and b) bending rigidity \citep{dai2016structure} of the constituent 2D materials. Second, our model does not predict spontaneous nonzero polarization in the absence of the applied field, nor does it predict hysteresis. While experiments \cite{yasuda2021stacking} report a net dipole moment under zero electric field, its origin remains unexplored. While the effect of domain-wall pinning has not been explored in the context of bilayer 2D materials, \citet{gao2011revealing} demonstrated that in bulk ferroelectric PbZr${_0.2}$Ti${_0.8}$O$_3$, the local disorder created by an isolated defect can significantly alter ferroelectric switching dynamics by modifying domain-wall velocities and locally favoring one polarization state over another. These results indicate that even weak, localized energy barriers introduced by individual defects can give rise to history-dependent switching behavior. Capturing such effects in heterodeformed bilayer hBN would require the explicit inclusion of defect-induced local energy barriers in the energy landscape, which is beyond the scope of the present work and will be explored in future studies.
Beyond these limitations, a natural extension of this work is to investigate in-plane ferroelectricity in heterodeformed bilayer hBN \cite{bennett2023polar}, which can be incorporated into the current model through an in-plane polarization landscape functional. We plan to explore this in future work. 

\section{Authors contributions}
\noindent
\textbf{Md Tusher Ahmed:} Conceptualization, Methodology, Software, Validation, Formal analysis, Investigation, Data Curation, Writing-Original Draft, Visualization,
\textbf{Chenhaoyue Wang:} Software, Formal analysis, Investigation, Data Curation, Writing-Original Draft,
\textbf{Amartya S. Banerjee:} Software, Investigation, Resources, Writing-Review \& Editing, Supervision, Funding acquisition
\textbf{Nikhil Chandra Admal:} Conceptualization, Software, Investigation, Resources, Writing-Review \& Editing, Supervision, Project Management, Funding acquisition.
\section{Conflicts of interest}
There are no conflicts to declare.
\section{Data availability}
The data supporting this article have been included in the Supplementary Information. The datasets generated during and/or analysed during the current study are available from the corresponding author on reasonable request.
\section{Acknowledgements}
NCA and TA would like to acknowledge support from the National Science Foundation Grant NSF-MOMS-2239734 with S. Qidwai as the program manager. ASB and CW would like to acknowledge support through grant DE-SC0023432 funded by the U.S. Department of Energy, Office of Science and through the UC National Laboratory Fees Research Program of the University of California, Grant Number L25CR9003. ASB and CW also acknowledge computational resource support from UCLA's Institute for Digital Research and Education (IDRE), and the National Energy Research Scientific Computing Center (NERSC awards BES-ERCAP0033206, BES-ERCAP0028072,  BES-ERCAP0025168 and BES-ERCAP0025205), a DOE Office of Science User Facility supported by the Office of Science of the U.S. Department of Energy under Contract No. DE-AC02-05CH11231.


\balance


\bibliography{rsc} 
\bibliographystyle{rsc} 
\end{document}